\documentclass[preprint2]{aastex}
\usepackage[dvips]{color} 

\newcommand{\myemail}{mawatari@astr.tohoku.ac.jp}

\shorttitle{CHARACTERIZATION OF THE 53W002 FIELD}
\shortauthors{Mawatari et al.}

\begin{document}


\title{CHARACTERIZATION OF THE DISTRIBUTION OF THE Ly{$\alpha$} EMITTERS IN THE 53W002 FIELD AT {\it z} = 2.4\altaffilmark{1}}


\author{K. Mawatari, T. Yamada and Y. Nakamura}
\affil{Astronomical Institute, Tohoku University, Aoba, Aramaki, Aoba-ku, Sendai, Miyagi, 980-8578, Japan}
\email{\myemail}
\author{T. Hayashino}
\affil{Research Center for Neutrino Science, Graduate School of Science, Tohoku University, Sendai 980-8578, Japan}
\and
\author{Y. Matsuda}
\affil{National Astronomical Observatory of Japan, Mitaka, Tokyo 181-8588, Japan}


\altaffiltext{1}{Based on the observations conducted with the Subaru Telescope which is operated by the National Astronomical Observatory of Japan.}


\begin{abstract}
We present the results of our wide-field narrow band imaging of the field around the radio galaxy \objectname{53W002} at $z$ = 2.390 with Subaru/Suprime-Cam. 
A custom made filter $NB413$ centered at 4140 \AA\ with the width of 83 \AA\ is used to observe the 31$'$ $\times$ 24$'$ area around the radio galaxy. 
We detected 204 Ly$\alpha$ emitters (LAEs) at $z = 2.4$ with a rest frame equivalent width larger than 25 \AA\ to the depth of 26 AB mag (in $NB413$). 
The entire LAE population in the 53W002 field has an average number density and distributions of equivalent width and size that are similar to those of other fields at $z \sim 2$. 
We identify a significant high density region (53W002F-HDR) that spreads over $\approx 5' \times 4'$ near 53W002 where the LAE number density is nearly four times as large as the average of the entire field. 
Using the probability distribution function of density fluctuation, we evaluate the rareness probability of the 53W002F-HDR to be 0.9$^{+2.4}_{-0.62}$\%, which corresponds to a moderately rich structure. 
No notable environmental dependency at the comoving scale of 10 Mpc is found for the distributions of the Ly$\alpha$ equivalent width and luminosity in the field. 
We also detected 4 Ly$\alpha$ blobs (LABs), one of which is newly discovered. 
They are all found to be located in the rims of high density regions. 
The biased location and unique morphologies in Ly$\alpha$ suggest that galaxy interaction play a key role in their formation. 

\end{abstract}





\section{INTRODUCTION}
In the local universe, a strong environmental dependence of galaxy properties is observed \citep[e.g.][]{dressler80,binggeli88,goto03,park07}. It is understood that external effects from their surrounding environments, namely galaxy-galaxy merger, harassment, ram-pressure stripping, and starvation may play key roles in sufficiently mature structures like clusters of galaxies in the nearby universe. At high redshift, on the other hand, the intrinsic environmental effects, such as galaxy formation bias \citep{Benson01,Weinberg04} should also play an important role rather than the external effects. The high density regions at high redshift are dominated by star-forming galaxies as galaxy formation preferentially occurs in dense environments at such epochs, which is naturally expected in the hierarchical structure formation theories. Observations of high-$z$ protoclusters are important in order to study how such intrinsic environmental effects work on the galaxies in the early universe. 

So far the narrow band method has been very efficient to discover density excesses at high redshift. Many authors have searched and discovered overdense structures traced by Ly$\alpha$ emitters (LAEs) since the late 1990s (e.g., {\it z} $\sim$ 2: \citealt{pentericci00,prescott08}, {\it z} $\sim$ 3: \citealt{steidel00,hayashino04,yamada12}, {\it z} $\sim$ 4: \citealt{venemans02,kuiper11}, {\it z} $\sim$ 5: \citealt{shimasaku03,venemans04}, {\it z} $\sim$ 6: \citealt{ouchi05}). LAEs are thought to be less massive, metal-poor, young and star-forming galaxies in general \citep{MalhotraRhoads02} although it is reported recently that their properties may change along redshift \citep{nilsson09,nakajima11}. \citet{ono10} repoted the typical physical quantities of LAEs at z = 3 - 4: M$_{star} = 10^{8} - 10^{8.5}$ M$_{\sun}$, SFR = 1 - 100 M$_{\sun}$ $yr^{-1}$, ages = $10^{6}$ - $10^{9}$ yr, E(B-V) $<$ 0.2. Such a galaxy population at high redshift is considered to trace the large scale structure \citep{uchimoto12}, and at least some fraction of them should be the building-blocks that evolve to massive galaxies at the later epochs.

 The distributions of Ly$\alpha$ emitters in the field around the radio galaxy \objectname{53W002} at $z$ = 2.39 and $\alpha = 17^{h}14^{m}14^{s}.8$, $\delta = +50^{\circ} 15' 49''$ (hereafter, the 53W002 field), which is the region we study in this paper, was first investigated by \citet{P96a,P96b}. They conducted deep medium-band ($\Delta\lambda$ = 150 \AA) imaging of the 53W002 field using {\it Hubble Space Telescope (HST)}/Wide Field Planetary Camera 2 (WFPC2). In their relatively small survey area ($\sim$ 2.5$'$ $\times$ 2.5$'$), 18 LAEs at {\it z} $\sim$ 2.4 were discovered, which was claimed to be the first direct detection of so-called `building-blocks' of galaxies at high redshift. \citet{P98} also observed three other WFPC2 fields using the same medium-band filter and argued that the 53W002 field is a high density region. \citet{K99} then surveyed the wider area around \objectname{53W002} with the 4 m Mayall telescope of Kitt Peak National Observatory. They found 14 LAEs in the area of 14$'$ $\times$ 14$'$ and suggested that the LAEs in the 53W002 field showed density excess compared with those in the blank fields. While several authors have investigated the properties of the galaxies in this field \citep{windhorst98,waddington98,yamada01,motohara01a,motohara01b,im02,keel02,smail03}, the previous observations seem to be insufficient to establish the significance of the density enhancement of the field. First, the medium-band filters used to detect Ly$\alpha$ emission line in the previous observations covers the relatively wide redshift range of $\Delta z$ = 0.12 ($z$ = 2.30 - 2.42), which may suffer from some projection effects. Second, both deep and wide observations are essential to identify the high redshift overdense region while the previous observations are not sufficient either in area ($HST$ observation) or in depth (ground-based observation). \\
 \indent
It is important not only to detect the high density regions at high redshift but also to characterize their richness in order to trace the evolution of galaxies in such dense environments or to compare the properties of the galaxies in their appropriate descendants. 
It is needed to introduce some objective measure to characterize such high-density overdense regions selected by LAEs to make the ancestors/descendants comparison possible. \\
\indent 
In this paper, we present the results of our deep and wide observations using the narrow-band filter.
We describe the observation and data reduction in $\S$2, sample selection of LAEs in $\S$3, and the results in $\S$4. In $\S$5, we discuss the implications of our results. Our conclusions are summarized in $\S$6. Throughout this paper, we use AB magnitude system and adopt a cosmology with $H_{0}=70.5$ km s$^{-1}$ Mpc$^{-1}$, $\Omega_{M}=0.27$, and $\Omega_{\Lambda}=0.73$ \citep{komatsu09}.

\section{OBSERVATION AND REDUCTION}

\begin{figure} [t]
\epsscale{1.00}
\plotone{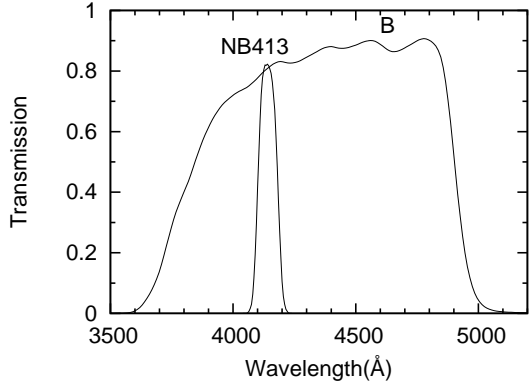}
\caption{Transmission curves of the filters we used: Subaru/Suprime-Cam $B$-band and the custom-made $NB413$.\label{fig1}}
\end{figure}

In 2009 May 22 - 23 (UT), we carried out imaging observations of the 53W002 field, centered at $\alpha=17^{h}14^{m}15^{s}.2, \delta=+50{\degr}15'49''$ (J2000) using Suprime-Cam \citep{miyazaki02} equipped on the 8.2 m Subaru telescope. 
We used the custom narrow-band filter $NB413$ which has the central wavelength of 4140 \AA\ and the width of 83 \AA\ ({\it FWHM}). 
The $NB413$ filter was built aiming to detect Ly$\alpha$ emission at $z$ = 2.37 - 2.44. 
We also used the $B$-band filter to estimate the continuum flux. 
In Figure~\ref{fig1} we show the transmission curves of the filters used in this study. 
The total exposure time was 3.5 hours for the $NB413$ and 1 hour for the $B$ band, respectively.\\
\indent
We used SDFRED version 2.0 \citep{yagi02,ouchi04} for the data reduction. 
As the number of the exposures is small (7 shots for the NB413 images), we made careful cosmic ray subtraction using L.A.Cosmic \citep{dokkum01} after the flat-fielding. 
We also masked the areas where bright stars affect the images. 
The final mosaiced images have the field of view (FoV) of $31.3'$ ${\times}$ $23.7'$ and the available area of 707.1 arcmin$^{2}$. 
This roughly corresponds to 50 Mpc $\times$ 40 Mpc in comoving scale. 
The width of $NB413$ filter corresponds to 90 Mpc. 
Our survey comoving volume for LAEs at {\it z} = 2.4 is 1.75 $\times$ 10$^5$ Mpc$^{3}$. 
The $B$ band image was smoothed so that the size of the stellar images is matched to that of the $NB413$ image; $0.86''$ in {\it FWHM}. 
The flux calibration was made by using the spectroscopic standard star, \objectname{PG1708+602}. 
We used SExtractor \citep{BerArn96} version 2.5.0 for object detection and photometry. 
The resultant 5$\sigma$ limiting magnitude in $2''$-diameter aperture is 25.95 mag for the $NB413$ image and 26.7 mag for the $B$ image. 
The 5$\sigma$ limiting magnitude of the $NB413$ image corresponds to the narrow-band luminosity (L$_{NB413}$) of 1.0 $\times$ 10$^{42}$ erg s$^{-1}$ assuming objects at {\it z} = 2.4.

\section{SELECTION OF LAE}
Object detection was made on the $NB413$ image: we adopted the criteria that more than 12 contiguous pixels above the 3$\sigma$ threshold of the background fluctuation. 
Furthermore, we rejected the obvious false detections using one of the SExtractor parameters, FLAGS. 
As a result, a total of $\sim$ 28,000 objects were detected in the range of 17 mag $\leq$ $NB413$ $\leq$ 27 mag. 
In the photometry, we used the fixed diameter (2$''$) aperture for both of the $NB413$ and the $B$ images, where each aperture was centered on the position of the object detected in the $NB413$ image.
For objects fainter than 1 $\sigma$ limiting magnitude in the $B$ band image, we replaced their $B$ band magnitude with 1 $\sigma$ (28.45 mag).\\
\begin{figure}[t]
\epsscale{1.00}
\plotone{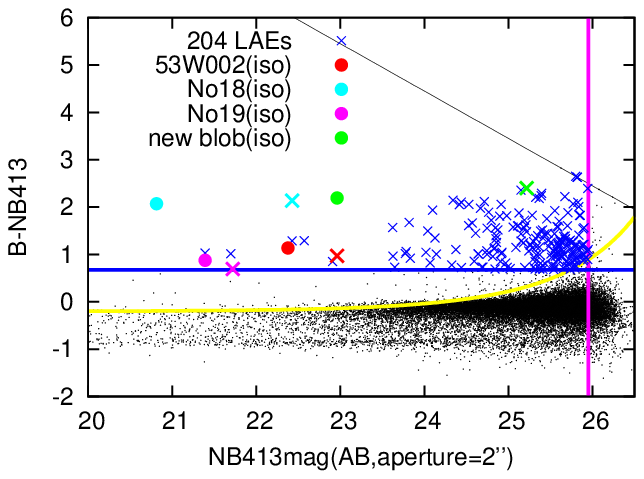}
\caption{The $B-NB413$ versus $NB413$ color magnitude diagram for detected objects in the 53W002 field. The colored lines show the LAE selection criteria: $B-NB413=0.68$ ({\it blue line}), the $5\sigma$ detection limit of the NB413 image ({\it magenta line}), and the $4\sigma$ color error ({\it yellow curve}). 
Colored crosses show the aperture colors of the selected LAEs; all ({\it blue}), 53W002 ({\it red}), No.18 ({\it cyan}), No.19 ({\it magenta}), and LAB4 ({\it green}). 
Filled circles show the isophotal colors for the blobs only, where correspondence between the object and the mark's color is the same as that of the crosses. 
The upper bound of the color ({\it black line}) corresponds to the $1\sigma$ detection limit of the $B$ image.\label{fig2}}
\end{figure}
\indent
To select the LAE candidates, we adopted the following three selection criteria. 
(i) $B-NB413$ $\geq$ 0.68: the $B-NB413$ color of 0.68 corresponds to the rest-frame equivalent width ({\it EW$_0$}) of 25 \AA\ for {\it z} = 2.4 Ly$\alpha$ emission line. 
(ii) 18 mag $\leq$ $NB413$ $\leq$ 25.95 mag. 
(iii) $B - NB413$ $\geq$ -0.2 + 4$\sigma$ in the color. 
The last criterion is to avoid the contamination at the faint end, where the foreground or background objects distribute around $B - NB413$ = -0.2. 
For the 213 objects which satisfy the three criteria, we visually inspected them to reject 9 sources (e.g. saturated stars, part of extended blobs). 
We finally accepted the remaining 204 objects as significant emission-line objects. 
These 204 candidates are shown in the color-magnitude diagram of Figure~\ref{fig2}. \\
\indent
One possible contaminant for {\it z} = 2.4 LAE is an [\ion{O}{2}] emitter at {\it z} = 0.10 - 0.12. 
Our 5$\sigma$ detection limit of the $NB413$ image corresponds to the [\ion{O}{2}] luminosity of L = 6.8 $\times$ 10$^{38}$ erg s$^{-1}$ and the surveyed comoving volume is $\sim$ 1,150 Mpc$^3$, which is one hundred times smaller than that for Ly$\alpha$ at {\it z} = 2.4. 
From the luminosity function and $EW$ distribution in \citet{sullivan00} we expect $\sim$ 7 [\ion{O}{2}] emitters in our sample. 
While another possible contaminant is a \ion{C}{4} emitter at {\it z} = 1.65 - 1.70, a strong \ion{C}{4} emission line is expected only among AGN whose number density is much smaller than [\ion{O}{2}] emitters or Ly$\alpha$ emitters. 
Considering the above, contamination can be negligible in our sample. 
Indeed, the previous spectroscopic observations \citep{matsuda05,matsuda06} show that such contamination in the similar sample at $z$ = 3.1 is at most 1\%. 
We ignore the contamination and treat all the 204 objects as LAEs at $z$ = 2.4 in the following discussion. \\
\indent
Table \ref{tb1} shows the properties of these LAEs. 
We listed not only the results of the fixed aperture photometry but also that of the Kron aperture photometry (SExtractor MAG\_AUTO) as many LAEs have the Ly$\alpha$ emission which is more extended than the continuum. 
Furthermore, we also performed the isophotal photometry for some extended Ly$\alpha$ clouds (i.e. Ly$\alpha$ blobs: LABs), and their isophotal colors are superposed in Figure~\ref{fig2} ({\it filled circle}). 
LABs will be discussed in detail in section~\ref{specials}. 
\begin{figure*}[t]
\begin{center}
\includegraphics[width=16cm]{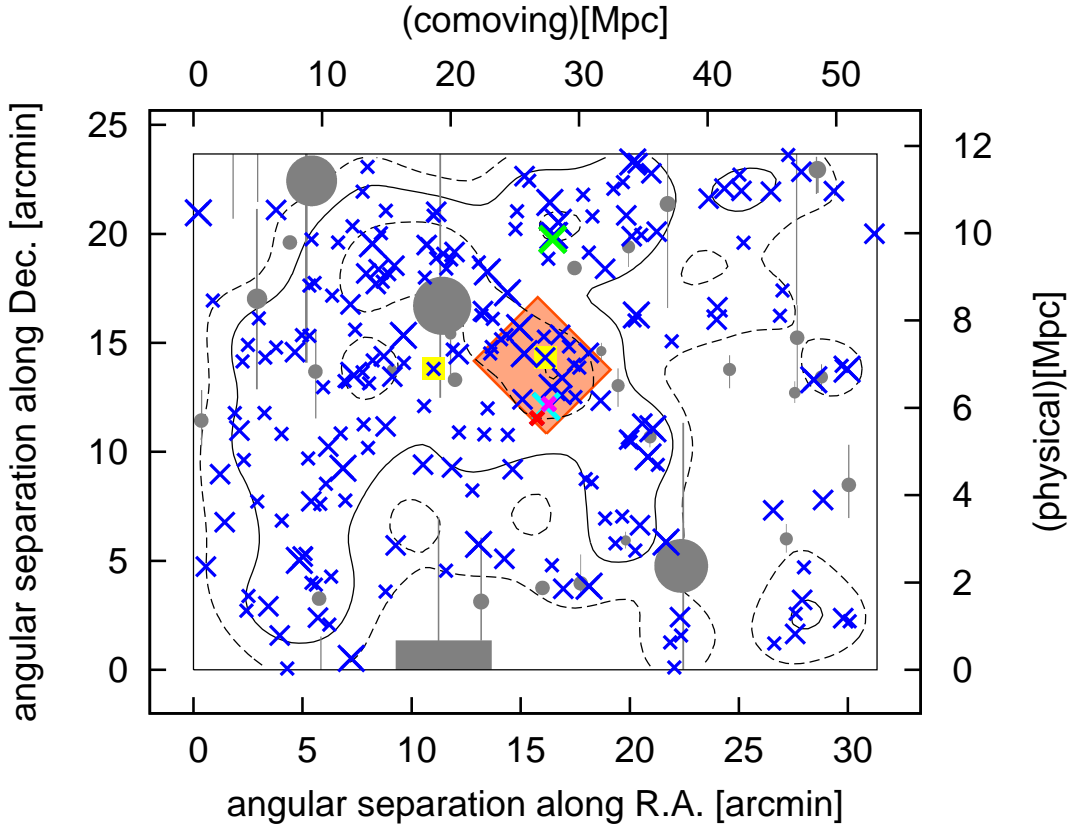}
\caption{Sky distribution of the 204 LAEs in the 53W002 field ({\it blue crosses}), which include extended Ly$\alpha$ Blobs (53W002: {\it red cross}, No.18: {\it cyan cross}, No.19: {\it magenta cross}, and LAB4: {\it green cross}) and QSOs that \citet{K99} reported ({\it yellow squares}). East is to the left and north is upward. {\it Grey shaded regions} correspond to the masked areas. {\it Black lines} correspond to contours of smoothed LAE number density: 0.5 $\times$, 1 $\times$ ({\it solid line}), 2 $\times$, and 2.85 $\times$ average of the entire field. We call the highest density region as 53W002F-HDR ({\it orange rectangle}).\label{fig3}}
\end{center}
\end{figure*}

\section{RESULTS}
\subsection{\it z = 2.4 Galaxies In The Entire Field}
\subsubsection{\it Spatial Distribution}
\begin{figure}[t]
\includegraphics[scale=0.55,angle=270]{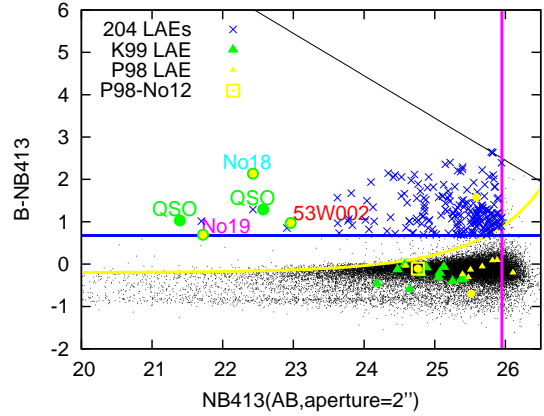}
\caption{The objects previously regarded as LAEs in the $B-NB413$ versus $NB413$ color magnitude diagram. The {\it yellow} and {\it green filled triangles} represent LAEs in \citet{P98} and \citet{K99}, respectively. LAEs that were spectroscopically confirmed as z $\sim$ 2.4 are shown by the {\it filled circles}. The {\it yellow open square} represents P98-No.12 that was spectroscopically confirmed as {\it z} = 2.388 in \citet{P96b,P98}.\label{fig4}}
\end{figure}

The sky distributions of the LAEs in the 53W002 field are shown in Figure~\ref{fig3}. 
The sizes of the symbols represent the rest-frame equivalent width at {\it z} = 2.4 based on the fixed aperture photometry (large: $EW_{0}$ $\geq$ 200 \AA, middle: 70 \AA\ $\leq$ $EW_{0}$ $\leq$ 200 \AA, small: 25 \AA\ $\leq$ $EW_{0}$ $\leq$ 70 \AA). 
The contours ({\it black line}) show the local surface number density obtained after being smoothed by a Gaussian kernel with $\sigma=1^{'}.5$ (0.5 $\times$, 1 $\times$, 2 $\times$, and 2.85 $\times$ average value of the entire field are shown). 
The mean number density of LAEs in the entire field is 0.289 arcmin$^{-2}$ or 1.16 $\times$ 10$^{-3}$ Mpc$^{-3}$. \\
\indent
First, we compare our results with the previous LAE search in this field \citep{P96a,P96b,P98,K99}. 
Figure~\ref{fig4} shows the objects selected as LAEs in \citet{P96b,P98} and \citet{K99} in our $B-NB413$ versus  $NB413$ color magnitude diagram. 
While the brightest objects and one of the LAEs are recovered in our sample, other objects that have been regarded as LAEs in previous studies are not. 
Their $B-NB413$ colors are distributed around $\sim 0$, which indicates that there is no significant emission line in the narrow-band. 
This result may be due to the difference of the filters. 
F410M used in \citet{P96a,P96b,P98} and F413M used in \citet{K99} are medium band filters that have the width of 150 \AA\ ($FWHM$) in the transmission curve, which samples the LAEs at {\it z} = 2.30 - 2.42 and 2.32 - 2.45, respectively. 
They are about twice as wide as our $NB413$ filter. 
Among the previously reported LAEs, 10 objects were spectroscopically confirmed as z $\sim$ 2.4 ({\it filled circles} in Figure~\ref{fig4}). 3 AGNs (53W002, No.18 and No.19), 2 QSOs in \citet{K99} and 1 LAE in \citet{P98} (No.29) are also selected in our analysis. No.11 in \citet{P98} is not selected as a LAE, which is naturally understood because its redshift of 2.45 is out of the range of our $NB413$. No.6 in \citet{P98} and No.5 in \citet{K99} locate near bright stars and we masked the field. 
No.12 in \citet{P96b,P98} is not selected as a LAE in our analysis while Ly$\alpha$ emission at its redshift of 2.388 should enter the $NB413$ filter (see {\it yellow open square} in Figure~\ref{fig4}). 
One possible explanation is that the claimed emission may be a transient such as supernovae at {\it z} = 2.4, but the reason  remains to be unknown. 

\subsubsection{\it Number Density of LAEs}\label{numden}
\setcounter{table}{1}
\begin{deluxetable}{ccccccc}
\tabletypesize{\footnotesize}
\tablecaption{Summary of the comparison of LAE number densities.\label{tb2}}
\tablewidth{0pt}
\tablehead{
\colhead{Reference} & \colhead{Field} & \colhead{$z$} & \colhead{Original criteria\tablenotemark{a}} & \colhead{Matched criteria\tablenotemark{b}} & \colhead{$\bar{n}$(field)\tablenotemark{c}} & \colhead{$\bar{n}$(53W002F)\tablenotemark{d}} \\
\colhead{} & \colhead{} & \colhead{} & \colhead{(L: 10$^{42}$ erg s$^{-1}$)} & \colhead{(L: 10$^{42}$ erg s$^{-1}$)} & \colhead{(10$^{-4}$ Mpc$^{-3}$)} & \colhead{(10$^{-4}$ Mpc$^{-3}$)} \\
\colhead{} & \colhead{} & \colhead{} & \colhead{($EW_0$: \AA)} & \colhead{($EW_0$: \AA)} & \colhead{} & \colhead{}
}
\startdata
\citet{nilsson09} & COSMOS & 2.25 & L$_{NB}$ $\ge$ 2.7 & L$_{NB}$ $\ge$ 2.7 & 4.8$\pm$0.3 & 3.0$\pm$0.4\\
 & & & $EW_0$ $\ge$ 20 & $EW_0$ $\ge$ 25 & & \\
\citet{stiavelli01} & DMS 2139-0356 & 2.4 & L$_{Ly\alpha}$ $\ge$ 2.4 & L$_{Ly\alpha}$ $\ge$ 2.4 & 1.3$\pm$0.2 & 2.1$\pm$0.3\\
 & & & $EW_0$ $\ge$ 25 & $EW_0$ $\ge$ 25 & & \\
\citet{prescott08} & LABd05 & 2.7 & L$_{NB}$ $\ge$ 1.7 & L$_{NB}$ $\ge$ 1.7 & 19$\pm$0.7 & 4.6$\pm$0.5\\
 & & & $EW_0$ $\ge$ 40 & $EW_0$ $\ge$ 40 & & \\
\citet{guaita10} & ECDF-S & 2.1 & L$_{Ly\alpha}$ $\ge$ 0.64 & ... & ...\tablenotemark{e} & ...\\
 & & & $EW_0$ $\ge$ 20 & ... & & \\
\enddata
\tablenotetext{a}{Original selection criteria of LAEs in individual studies. Note that they are expressed in our adopted cosmology.}
\tablenotetext{b}{Matched LAE selection criteria for the comparison with the number density for the 53W002 field.}
\tablenotetext{c}{Volume number densities of LAEs with matched selection criteria for individual fields.}
\tablenotetext{d}{Volume number densities of LAEs with matched selection criteria for the 53W002 fields.}
\tablenotetext{e}{It is mentioned that the LAE number density for ECDF-S is roughly the same with that of the $z$ = 2.25 LAEs in the COSMOS field.}
\end{deluxetable}
In the last section we obtained the LAE average number density in our survey field as 1.16 $\times$ 10$^{-3}$ Mpc$^{-3}$. 
This value is estimated for the LAEs with L$_{NB}$ brighter than 1.0 $\times$ 10$^{42}$ erg s$^{-1}$ and $EW_{0}$ larger than 25 \AA\ without correction of the IGM attenuation. 
We compare the LAE density of the 53W002 field with results of several other {\it z} $\sim$ 2 LAE studies. 
Note that the selection criteria as well as the adopted cosmological parameters are matched for the purpose, which is not always straightforward. 
\citet{nilsson09} surveyed the COSMOS field, and obtained a volume density of 6.2 $\times$ 10$^{-4}$ Mpc$^{-3}$ for {\it z} = 2.25 LAEs with L$_{NB}$ $\ge$ 2.3 $\times$ 10$^{42}$ erg s$^{-1}$ and $EW_{0}$ $\ge$ 20 \AA\ (in their adopted cosmological parameters). 
After correcting for the cosmological parameters, the number densities of LAEs with L$_{NB}$ $\ge$ 2.7 $\times$ 10$^{42}$ and $EW_{0}$ $\ge$ 25 \AA\ are (3.0$\pm$0.4) $\times$ 10$^{-4}$ Mpc$^{-3}$ and (4.8$\pm$0.3) $\times$ 10$^{-4}$ Mpc$^{-3}$ for the 53W002 field and the COSMOS field, respectively. 
In the same way we made a comparison with the density of LAEs with L$_{Ly\alpha}$ $\ge$ 2.4 $\times$ 10$^{42}$ erg s$^{-1}$ and $EW_0$ $\ge$ 25 \AA\ in \citet{stiavelli01}. 
They searched LAEs at {\it z} = 2.4 in the field around QSO DMS 2139-0356 and showed the volume density of (1.3$\pm$0.2) $\times$ 10$^{-4}$ Mpc$^{-3}$. 
With the same limit, the volume density for the 53W002 field is estimated as (2.1$\pm$0.3) $\times$ 10$^{-4}$ Mpc$^{-3}$. 
\citet{prescott08} studied the overdense environment around a large LAB and found 785 LAEs at {\it z} = 2.7, which results in the volume density of (1.9$\pm$0.07) $\times$ 10$^{-3}$ Mpc$^{-3}$ for the threshold of L$_{NB}$ $\ge$ 1.7 $\times$ 10$^{42}$ erg s$^{-1}$ and $EW_0$ $\ge$ 40 \AA. 
The density for the 53W002 field is estimated as (4.6$\pm$0.5) $\times$ 10$^{-4}$ Mpc$^{-3}$ for the same threshold. 
The {\it z} = 2.7 fields clearly has the larger density than our {\it z} = 2.4 fields. 
\citet{guaita10} surveyed the Extended Chandra Deep Field-South (ECDF-S) and estimated the density of the {\it z} = 2.1 LAEs as (3.1$\pm$0.9) $\times$ 10$^{-3}$ Mpc$^{-3}$. 
We cannot make a direct comparison as their LAE selection threshold is deeper than ours and there is no catalog of their LAEs. 
The LAE density for ECDF-S may be comparable or slightly larger than that for the 53W002 field because they mentioned that the density of ECDF-S is roughly the same with that of the {\it z} = 2.25 LAEs in the COSMOS field \citep{nilsson09}. 
Table \ref{tb2} summarizes the comparison of the LAE number density of the 53W002 field with that of the four other $z$ $\sim$ 2 fields. 
These comparisons suggest that the LAE volume density in the entire 53W002 field is comparable to those in other {\it z} $\sim$ 2 general fields. \\

\subsubsection{\it Sizes and EWs of LAEs}\label{size-ew}
\begin{figure}[t] 
\includegraphics[scale=0.6,angle=270]{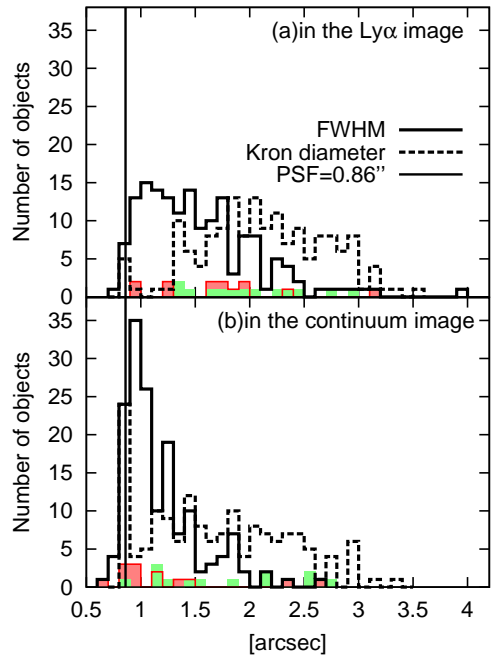}
\caption{The size distribution of the LAEs in the 53W002 field. {\it Top}: The size in the Ly$\alpha$ image. {\it Bottom}: The size in the continuum image. In each panel, the {\it solid black histogram} represents the $FWHM$ distribution of the objects in the entire field, and {\it red shaded histogram} represents that in 53W002F-HDR. On the other hand {\it dashed black histogram} and {\it green shaded histogram} represents the Kron diameter distribution of the objects in the entire field and 53W002F-HDR, respectively. We also show the width of the point sources in $FWHM$, 0.86$''$ ({\it horizontal solid line})\label{fig5}.}
\end{figure}

In order to evaluate the size of the Ly$\alpha$ emission, we first made the Ly$\alpha$ image of the LAEs by subtracting the continuum component from the $NB413$ image. 
Assuming the flat spectrum of the continuum in the range of $B$-band, we separated the Ly$\alpha$ component from the continuum component as
\begin{mathletters}
\begin{eqnarray}
f_{{\nu}NB413} & = & f_{{\nu}C} + f_{{\nu}Ly\alpha} , \label{eq:test-fnuNB}\\
f_{{\nu}B} & = & f_{{\nu}C} + f_{{\nu}Ly\alpha}{\times}\frac{{\Delta}\nu_{NB413}}{{\Delta}\nu_{B}} ,\label{eq:test-fnuB}
\end{eqnarray}
\end{mathletters}
where $f_{{\nu}NB413}$ and $f_{{\nu}B}$ is the flux per unit frequency (erg s$^{-1}$ cm$^{-2}$ Hz$^{-1}$) for the photons that enter the $NB413$ and $B$-band filter, $f_{{\nu}Ly\alpha}$ and $f_{{\nu}C}$ is the flux per unit frequency of the Ly$\alpha$ line and the continuum component at the wavelength of 4140 \AA, and ${\Delta}\nu_{NB413}$ and ${\Delta}\nu_{B}$ is the width of the $NB413$ and $B$-band filter ($FWHM$) in unit of frequency, respectively. 
We also made the continuum image by subtracting the Ly$\alpha$ component from the $B$ image. 
The 5$\sigma$ limiting magnitude in a $2''$ diameter aperture is 26.0 mag for the Ly$\alpha$ image and 26.6 mag for the continuum image. 
Among the 204 LAEs we carefully selected the objects to measure the sizes which are not easily affected by background noises especially in the continuum image. 
We made object detection independently on the Ly$\alpha$ and continuum images, and selected the objects that have fluxes brighter than the 2$\sigma$ detection limit in both images (Ly$\alpha$ $<$ 27.0 mag, continuum $<$ 27.6 mag).  
This flux criterion is aimed to select objects that have peak higher than twice of the noise, which leads to the robustness in measuring $FWHM$s of the profiles. 
The number of objects in the resultant sample is 158, and we used SExtractor to mesure their sizes (FWHM\_IMAGE, KRON\_RADIUS). 
Figure~\ref{fig5} shows the size distribution in the Ly$\alpha$ image and the continuum image. 
The $FWHM$ distributions ({\it solid line}) suggest that the samples are more extended in Ly$\alpha$ than in continuum, which is a common trend for LAEs \citep{steidel11,matsuda12}. 
There are no notable differences between the $FWHM$ distribution in the 53W002 field and that in another {\it z} $\sim$ 2 general field reported in \citet{nilsson09}. 
We also show the Kron diameter distributions ({\it dashed line}), which are consistent with the fact that the fluxes using aperture of 2.5 $\times$ Kron diameter are systematically larger than those using the fixed aperture (see Table \ref{tb1}). \\
\begin{figure}[t]
\includegraphics[scale=1.0]{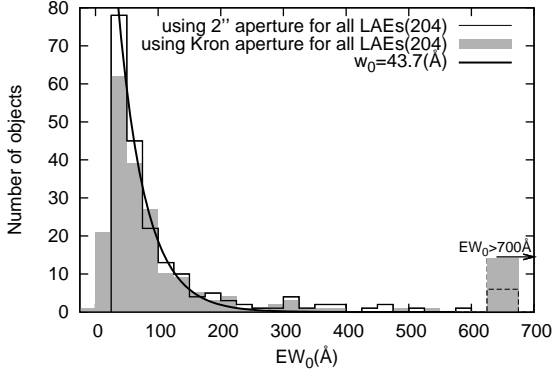}
\caption{The Ly$\alpha$ $EW_0$ distribution of the 204 LAEs in the 53W002 field. {\it Solid histogram} shows the $EW_0$ based on the fixed aperture (2$''$) photometry, and {\it grey shaded histogram} shows the $EW_0$ based on the Kron aperture photometry. Bin size is 25 \AA. {\it Solid curve} indicates the exponential fit to the 2$''$ aperture $EW_0$ distribution. As there are some objects out of the $EW_0$ range of the figure, we show them in the $EW_0$ = 650 \AA\ bin.\label{fig6}}
\end{figure}
\indent
For the $EW_0$ distribution, it was suggested in previous studies \citep{gronwall07,nilsson07,nilsson09,cassata11} that the fraction of LAEs with small $EW_0$ increase at lower redshift. 
For comparison, we show the distribution of $EW_0$ of our LAEs in the 53W002 field in Figure~\ref{fig6}. 
Note that these $EW_0$ are based on the simple calculation assuming neither Galactic nor Intergalactic medium (IGM) extinction. 
Following \citet{gronwall07}, we then fitted the $EW_0$ distribution calculated from the fixed aperture photometry with an exponentially declining function:
\begin{equation}\label{eq:2}
N = C \times e^{-EW_0/w_0},
\end{equation}
where $C$ is a normalization constant and $w_0$ is e-folding length. 
The best fit solution minimizing the $\chi^2$ is $w_0$ = 43.7 $\pm$ 0.43 \AA.
While this e-folding length is also slightly smaller than that of 48.5 $\pm$ 1.7 \AA\ for {\it z} = 2.25 LAEs \citep{nilsson09}, it is much smaller than that of 76 $^{+11}_{-8}$ \AA\ for {\it z} = 3.1 LAEs \citep{gronwall07}. 
It is consistent with the trend that $EW_0$ distribution becomes narrower at lower redshift. 
It was also pointed out by the previous studies that the number of LAEs with $EW_{0}$ $\ge$ 240 \AA\ is less than that at higher redshift ($z \gtrsim 4$), where such a extremely large $EW_0$ cannot be explained without top-heavy initial mass function or Population III stars \citep{MalhotraRhoads02}. 
The fraction of our LAEs that have $EW_0$ larger than 240 \AA\ is 0.11 $\pm$ 0.02 (23/204) either in the case of 2$''$ aperture, or the Kron aperture. 
This fraction for the 53W002 field is larger than the value, a few percent, for other fields at {\it z} $\sim$ 2 \citep{nilsson09,stiavelli01}. 
It may be unique property of the 53W002 field that there are substantial number of LAEs that have extremely large $EW_0$ while the rest distribution is dominated by small $EW_0$ objects. 
We also note that there is no notable trend about the sky distribution of high $EW$ objects ($EW_0$ $>$ 240 \AA).

\subsubsection{\it Blobs, AGNs and QSOs} \label{specials}
\begin{figure*}[p]
\begin{center}
\includegraphics[width=16cm]{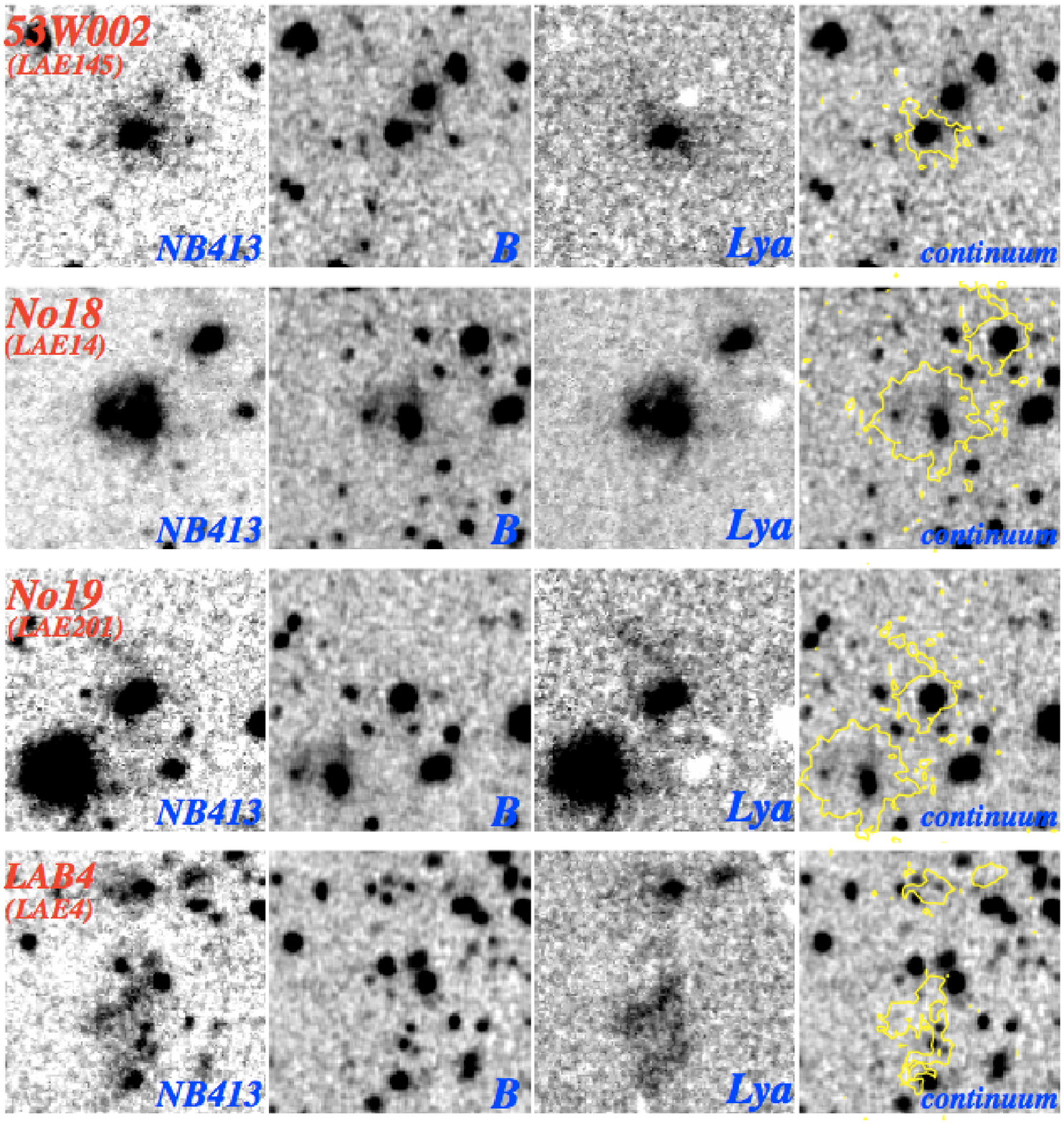}
\caption{$B$, $NB413$, Ly$\alpha$, and continuum images of the 4 LABs in the 53W002 field. North is up, and each panel's size is 24 $\times$ 24 arcsec$^2$ which corresponds to 200 $\times$ 200 kpc$^2$ in physical scale. {\it Yellow contours} superposed on continuum images show the isophotal apertures of the Ly$\alpha$ image, 2$\sigma$ arcsec$^{-2}$ of the background fluctuation (7.90 $\times$ 10$^{-18}$ erg s$^{-1}$ cm$^{-2}$ arcsec$^{-2}$). \label{fig7}}
\end{center}
\end{figure*}

\begin{deluxetable}{ccccccccll}
\tabletypesize{\scriptsize}
\rotate
\tablecaption{Properties of The LABs in The 53W002 Field\label{tb3}}
\tablewidth{0pt}
\tablehead{
\colhead{ID} & \colhead{R.A.} & \colhead{Dec.} & \colhead{Area\tablenotemark{a}} & \colhead{$a$\tablenotemark{b}} & \colhead{$EW_0$\tablenotemark{c}} & \colhead{log L$_{Ly\alpha}$\tablenotemark{c}} & \colhead{$\log\langle$SB$\rangle$\tablenotemark{d}} & \colhead{Note} & \colhead{Ref}\\
\colhead{}  & \colhead{(deg)} & \colhead{(deg)} & \colhead{(arcsec$^2$)} & \colhead{(kpc)} & \colhead{(iso, $\AA$)} & \colhead{(iso, erg s$^{-1}$)} & \colhead{(erg s$^{-1}$ cm$^{-2}$ arcsec$^{-2}$)} & \colhead{} &\\
\colhead{(1)} & \colhead{(2)} & \colhead{(3)} & \colhead{(4)}  & \colhead{(5)} & \colhead{(6)} & \colhead{(7)}  & \colhead{(8)}  & \colhead{(9)} & \colhead{(10)}
}
\startdata
53W002 & 258.56128 & 50.25853 & 16.32 & 59.3 & 59.3 & 43.294 & -16.592 & radio & 1\\
(LAE145) & & & & & & & & &\\
No.18 & 258.55016 & 50.26759 & 61.6 & 93.3 & 325 & 44.037 & -16.425 & AGN/QSO & 2,3\\
(LAE14) & & & & & & & & &\\
No.19 & 258.54727 & 50.26960 & 19.28 & 88.5 & 37.4 & 43.620 & -16.338 & QSO & 2\\
(LAE201) & & & & & & & & &\\
LAB4 & 258.54414 & 50.39581 & 25.8 & 90.4 & 439 & 43.185 & -16.899 & ... & ...\\
(LAE4) & & & & & & & & &\\
\enddata
\tablenotetext{a}{Isophotal area above 2$\sigma$ arcsec$^{-2}$ of the background fluctuation of the Ly$\alpha$ image.}
\tablenotetext{b}{Major-axis diameter of the isophotal aperture in physical scale at {\it z} = 2.4.}
\tablenotetext{c}{The sixth and seventh rows are based on the isophotal aperture photometry.}
\tablenotetext{d}{Average Ly$\alpha$ surface brightness.}
\tablerefs{
(1) Windhorst et al. 1991; (2) Pascarelle et al. 1996b; (3) Smail et al. 2003.}
\end{deluxetable}
There are some unique objects in the 53W002 field. 
We found four Ly$\alpha$ blobs (LABs), three of which are also reported in previous works. 
The LABs were selected according to a similar method described in \citet{matsuda04}. 
First, we made the object detection on the Ly$\alpha$ image adopting the criteria to have more than 20 contiguous pixels above the threshold of 2$\sigma$ of the background fluctuation, which corresponds to 27.1 mag arcsec$^{-2}$ (7.90 $\times$ 10$^{-18}$ erg s$^{-1}$ cm$^{-2}$ arcsec$^{-2}$). 
Next, the $NB413$ and $B$ magnitudes of the detected objects were measured using the same isophotal apertures defined in the process of source detection on the Ly$\alpha$ image. 
We then selected the LABs by adopting the following criteria. 
(i)$NB413 - B$(isophotal) $\ge$ 0.68 \& $NB413$(isophotal) $\leq$ 25.95: these are the criteria to select emitters. 
(ii)Isophotal area in the Ly$\alpha$ image $\ge$ 10 arcsec$^2$: the threshold of 10 arcsec$^2$ corresponds to the area of a circle with a diameter of 30 kpc in physical scale at {\it z} = 2.4. 
We refer to the detected four LABs as 53W002 (LAE145), No.18 (LAE14), No.19 (LAE201) which are named according to the previous works \citep{W91,P96b}, and LAB4 (LAE4) which was newly found in this study. \\
\indent
The properties of the four LABs (including the isophotal $B - NB413$ colors) are shown in Table \ref{tb3} and Figure~\ref{fig2} ({\it filled circles}). 
In Table \ref{tb3} we show not only the isophotal areas but also the major-axis diameters \citep{matsuda11}. 
The $NB413$, $B$, Ly$\alpha$ and continuum images of the four LABs are shown in Figure~\ref{fig7}, where the 2$\sigma$ isophotal apertures of the Ly$\alpha$ image that we used in both of the detection and photometry are superposed on the continuum images. 
\objectname{53W002}, which is originally  found as a weak radio source \citep{W91}, has the smallest isophotal area and major-axis among the four LABs. \\
\begin{figure*}[t]
\begin{center}
\includegraphics[width=16cm]{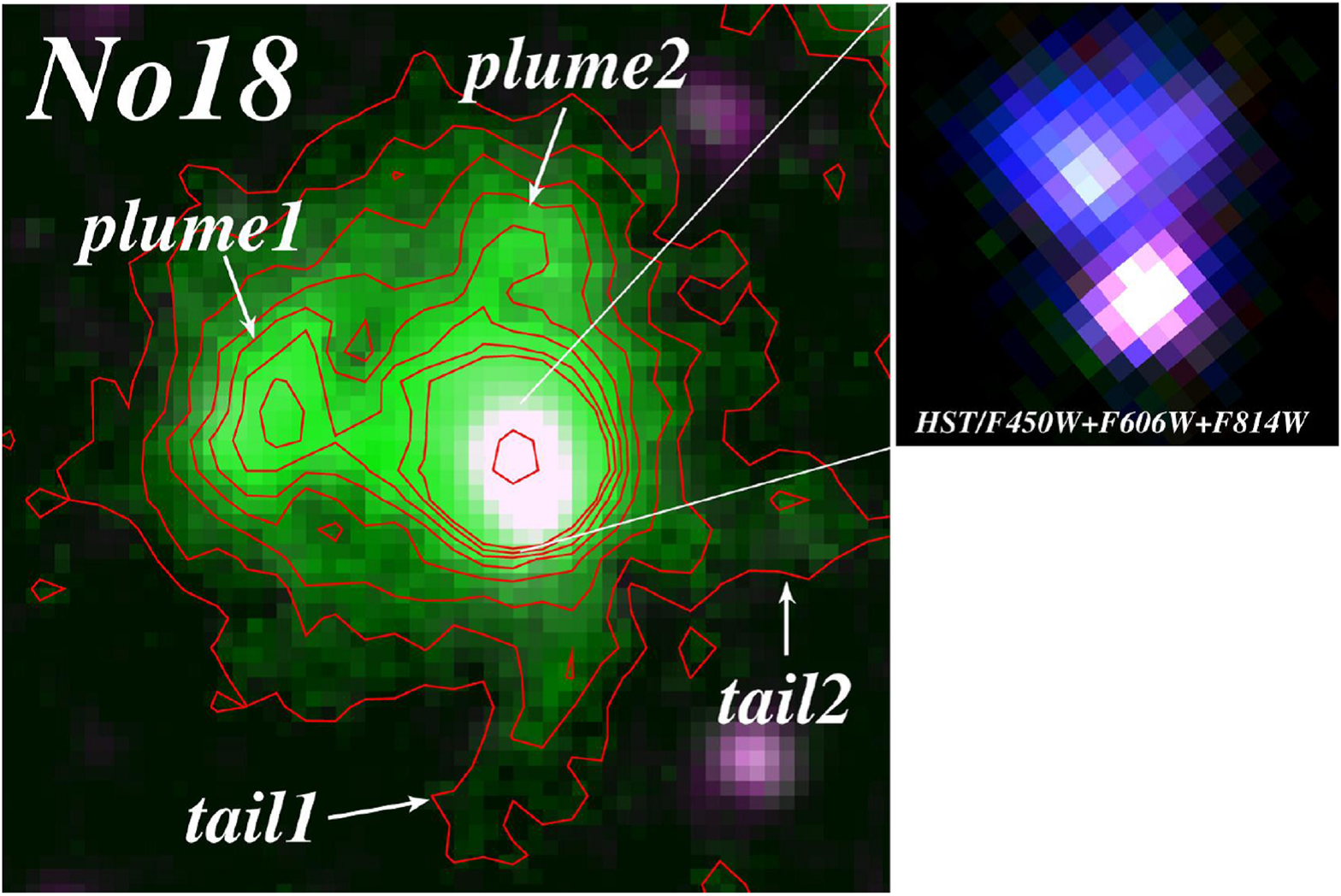}
\caption{Pseudo-color images of No.18 whose color consists of $NB413$ for green, and $B$ for blue and red (the left big panel). The panel's size is 12 $\times$ 12 arcsec$^2$ which corresponds to 100 $\times$ 100 kpc$^2$ in physical scale. {\it Red lines} superposed on the image correspond to the isophotal contours of the Ly$\alpha$ image at levels of 2, 4, 6, 8, 10, 12, 14, 16, and 100$\sigma$ arcsec$^{-2}$. The small right panel is the enlarged image of the central core. Colors are F450W for blue, F606W for green, and F814W for red. The panel's size is 2 $\times$ 2 arcsec$^2$ (physical 17 $\times$ 17 kpc$^2$). \label{fig8}}
\end{center}
\end{figure*}
\indent
No.18 lies in the northwest direction from 53W002, and it is the most unique object with several very interesting features. 
First it is notable that No.18 has an extremely large $EW_0$ and very extended Ly$\alpha$ nebulae in spite of a substantial amount of dust (see later description). 
We show the pseudo-color image of No.18 ($NB413$ for green, $B$ for blue and red) as well as their Ly$\alpha$ surface brightness in Figure~\ref{fig8}. 
It clearly has two bright structures in Ly$\alpha$ ($plume1$ and $plume2$). 
The unique shape of $plume1$ appears like a bow-shock front with a jet expelled from the core. 
A weak continuum source is seen in the Ly$\alpha$ isophotal contour of $plume1$, although there is a small offset between their positions . 
No.18 also has two tails with lower surface brightness ($tail1$ and $tail2$). 
They resemble tidal tails, which may indicate that No.18 is undergoing a merger process. 
On the other hand, $tail1$ and $tail2$ are seen only in the Ly$\alpha$ image and no significant continuum emission is detected. 
If these tails consist only of gas and not of stars the energy source of Ly$\alpha$ emission remains to be considered. 
In Figure~\ref{fig8} we also show the enlarged image of the central core that is unresolved in our continuum image but resolved in $HST$ images. 
Note that we used the $HST$ archive data (WFPC2/F450W, F606W, and F814W) to make the color image, which are the same as that used in the previous study \citep{P96b}. 
The core consists of two components: the compact southern component that has redder color in the rest-frame optical/UV, and the more diffuse northern component that even has bluer color and weaker continuum brightness, though a large fraction of the Ly$\alpha$ emission comes from this component \citep{smail03}. 
Near-infrared spectroscopy classified the southern component as a Seyfert 2 AGN \citep{motohara01a}. 
\citet{smail03} discovered the submillimeter counterpart (SMG) to No.18. 
They also summarized several studies about No.18 that its physical process may be the complex mix of star formation, AGN activity of the southern component, resulting local ionization or illumination of the northern component and extended superwind driven by highly obscured starburst. 
Our results are well consistent with their picture about the core as the northern component has a large fraction of Ly$\alpha$ (see the 100 $\sigma$ isophotal contour in Figure~\ref{fig8}). 
$Plume1$ appears to be expelled from the northern component, which is consistent with the appearance of the high spatial resolution Ly$\alpha$ image by $HST$ \citep[see][]{smail03}. 
Further observation, especially for $plume1$ and the northern component that have never been observed spectroscopically, is needed in order to get a definitive picture about the physical process occurred in No.18. \\
\indent
No.19, which lies close by No.18, was found as a variable QSO and spectroscopically confirmed to be at {\it z} = 2.397 \citep{P96a}. 
In the Ly$\alpha$ image it shows faint and clumpy tails that is detected for the first time. 
Including the tails we estimated the size as 88.5 kpc in physical scale, which is relatively large among the LABs in other fields. 
Unlike No.18, however, we consider that No.19 is physically dominated by AGN because of its small $EW_0$. \\
\indent
LAB4 (the south object of the bottom row in the stamps in Figure~\ref{fig7}) is located in the region previously not observed and with two blobby companions in the north. 
Unlike the other three objects associated with AGNs, LAB4 has no strong continuum counterpart and looks filamentary in Ly$\alpha$. \\
\indent
We also detected two QSOs previously reported \citep{K99} as LAEs (called as LAE77 and LAE128). 
LAE128 is located at the peak of the smoothed LAE number density in the 53W002 field, and LAE77 is also at relatively higher density region (see {\it yellow squares} in Figure~\ref{fig3}).

\subsection{\it 53W002F-HDR}
The most dense part in our surveyed field is at around ($\alpha$,$\delta$) = (17$^h$14$^m$10$^{s}$.0,+50$\degr$17$'$30$''$) where the local number density is $\sim$ 2.9 times of the average of the entire field. 
We identify the 4.7$'$ $\times$ 4.1$'$ region around the most dense point as the high density region and refer this as 53W002F-HDR (see the {\it orange rectangle} in Figure~\ref{fig3}). 
53W002F-HDR has a volume of 4850 Mpc$^{3}$ in comoving scale and the LAE total number of 21. 
So the mean number density of the 53W002F-HDR is 4.33 $\times$ 10$^{-3}$ Mpc$^{-3}$, which is nearly four times as large as the mean of the entire field. 
The four unique objects (53W002, No.18, No.19 and LAE128) lie in the very small volume of 53W002F-HDR. 
In Figure~\ref{fig5} the $FWHM$ ({\it red shaded}) and Kron diameter ({\it green shaded}) distributions of the LAEs in 53W002F-HDR are presented.

\section{DISCUSSION}\label{discussion}

\subsection{\it Rareness Probability of 53W002F-HDR}
\begin{figure}[t]
\includegraphics[scale=1.2]{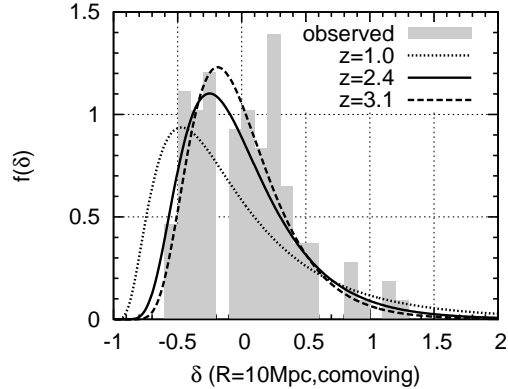}
\caption{The PDF of mass fluctuations with the scale of $R$ = 10 Mpc for three different cosmic epochs: {\it z} = 1.0 ({\it dotted curve}) simply for a comparison, {\it z} = 2.4 ({\it solid curve}) that is the epoch we observe in this study, and {\it z} = 3.1 ({\it dashed curve}) that is the epoch of the supplement data noticed in Section \ref{environment-dependence}. They are predicted theoretically from Eq.(\ref{eq:massPDF}). We also show the mass PDF obtained observationally from the data of LAEs at $z$ = 2.4 in the 53W002 field ({\it grey shaded histogram}), where we used the linear bias to convert the LAE overdensity to the mass overdensity (see text for details). A KS test does not reject the null hypothesis that the observed mass PDF is drawn from the theoretical PDF at $z$ = 2.4 (KS probability = 0.99).\label{fig9}}
\end{figure}

In this section we are going to discuss how significant 53W002F-HDR is. 
We should apply a more objective indicator of the significance of the structure than just density so that we can also make comparison with proto-cluster regions at other redshift and to trace the evolution of galaxies in such dense environment. 
We introduce a new quantity for this purpose. 
\\
\indent
Assuming the number density of the LAEs in the entire survey field as the average in the {\it z} = 2.4 universe ($\bar{n} = 1.16 \times 10^{-3}$ Mpc$^{-3}$), the number density of 53W002F-HDR ($n_{HDR} = 4.33 \times 10^{-3}$ Mpc$^{-3}$) is converted to the LAE overdensity,
\begin{equation}
\delta_{LAE,HDR} = \frac{n_{HDR}-\bar{n}}{\bar{n}} = 2.7 \pm 0.8,
\end{equation}
where the error is the poisson error. 
Although this represents the overdensity of the LAE number density, we assume this can be regarded as the LAE mass overdensity, $(\rho_{HDR}-\bar{\rho})/\bar{\rho}$. 
Furthermore we approximated the overdensity of the underlying mass assuming the LAE linear bias by
\begin{equation}
\delta_{LAE} = b_{LAE} \times \delta_{mass}.
\end{equation}
There is only a few studies about bias for {\it z} $\sim$ 2 LAEs and almost all of them estimated linear bias. 
So we used the linear bias of $b_{LAE} = 1.8 \pm 0.3$, which was estimated from the observation of 250 LAEs at {\it z} = 2.1 \citep{guaita10}. 
Thus, the mass overdensity of 53W002F-HDR results in $\delta_{mass,HDR} = 1.5 \pm 0.5$. \\
\indent
Then we introduce the probability distribution function (PDF) to characterize the general distribution of mass fluctuations in the specific redshift universe. 
By definition, the PDF of cosmological density fluctuations describes the probability of having a fluctuation in the overdensity range ($\delta,\delta+d\delta$), within a spherical region of characteristic radius $R$ randomly located in the universe at the given redshift. 
We followed mainly \citet{marinoni05,marinoni08} to make the PDF of the underlying mass theoretically. 
Its shape is well approximated by a lognormal distribution up to moderately non-linear regimes \citep{ColesJones91,kofman94,TaylorWatts00,kayo01},
\begin{equation}
f_{R}(\delta) = \frac{(2\pi\omega^{2}_{R})^{-1/2}}{1+\delta}exp\biggl[-\frac{\{\ln(1+\delta)+\omega^{2}_{R}/2\}^2}{2\omega^{2}_{R}}\biggr].\label{eq:massPDF}
\end{equation}
This formula is fully characterized by a single parameter $\omega_R$, which is related to the variance of the overdensity field on the scale $R$ ($\sigma^{2}_{R} \equiv \langle\delta^2\rangle_R$) as
\begin{equation}
\omega^{2}_{R} = \ln(1+\sigma^{2}_{R}).\label{eq:omega2}
\end{equation}
The linear growth theory approximation is used to estimate $\sigma_R(z)$ with the normalization factor of $\sigma_8$ = 0.81 \citep{komatsu11}. 
We applied the correction of the redshift distortion to $\sigma_R(z)$ in Eq.(\ref{eq:omega2}) because we need to make the PDF in redshift space in order to compare with the observations. 
The correction accounts for the average contribution of the linear redshift distortions induced by peculiar velocities \citep{kaiser87}. 
See \citet{marinoni05} for more details. 
Figure~\ref{fig9} shows the examples of the PDF of mass fluctuations for three different epochs, where we adopt the radius of sampling spheres of $R$ = 10 Mpc. 
We also attempt to reconstruct mass PDF from the observed LAE distribution. 
First, we measured the LAE overdensities within a circular regions with the radius of 2.3$'$ (comoving 3.9 Mpc) randomly located in the field except for the edge. 
These circular regions have the equivalent volume with spheres with radius of 10 Mpc.  
Next, we assume the linear bias ($b_{LAE}$ = 1.8) to convert the LAE overdensities to mass overdensities. 
The resulting mass PDF is shown in Figure~\ref{fig9} ({\it grey shaded histogram}). 
The theoretical and observed PDFs of mass fluctuation at $z$ = 2.4 seem to be in reasonably good agreement. 
A Kolmogorov-Smirnov (KS) test gives the significance probability of the difference between the two distributions (KS probability, ranging between 0 and 1), 0.99.\\
\indent
We evaluated the significance of 53W002F-HDR as follows. 
First we made the theoretical mass PDF at {\it z} = 2.4 with $R$ = 10.5 Mpc, this scale corresponds to the radius of a sphere having the equivalent volume with 53W002F-HDR in comoving scale. 
Next we calculated the probability of having $\delta_{mass}$ larger than that of 53W002F-HDR (1.5 $\pm$ 0.5) from the mass PDF, 0.9$^{+2.4}_{-0.62}$\%. 
This implies that 53W002F-HDR is a rich region in the {\it z} = 2.4 universe where such overdense structure exist with the probability of $\sim$ 0.9\%. 
Hereafter we call the probability representing rareness of structure as ``rareness probability''. \\
\indent
Next, we compare the rareness probability of 53W002F-HDR with those of other high density regions traced by LAEs at $z$ $>$ 2. 
For comparison, we used the data of the SSA22 field at z = 3.1 \citep{yamada12}, the TN J1338-1942 field at z = 4.1 \citep{venemans02}, and the 6C0140+326 field at z = 4.4 \citep{kuiper11}, where their comoving volumes and LAE overdensities are available. 
The central region in the SSA22 field (Sb1 peak) has comoving volume of 20,580 Mpc$^{3}$ and LAE overdensity of 3.36$\pm0.03$. 
We calculated the rareness probability of the Sb1 peak assuming LAE linear bias of 1.9 \citep{guaita10}, which results in the rareness probability of 0.0017$^{+0.0002}_{-0.0001}$\%. 
The region around the radio galaxy TN J1338-1942 has comoving volume of 7,315 Mpc$^3$ and LAE overdensity of 4$\pm1.4$, whose rareness probability is 0.2$^{+1.5}_{-0.17}$\% assuming the LAE linear bias of 3.7 \citep{kovac07}. 
The comoving volume and LAE overdensity of the region around the radio galaxy 6C0140+326 is 1570 Mpc$^{3}$ and 8$\pm5$, respectively. 
This corresponds to the rareness probability of 0.043$^{+3.49}_{-0.042}$\%, where we assume that the LAE linear bias is 3.7 \citep{kovac07}. 
Compared with these values, the rareness probability of 53W002F-HDR, 0.9\%, is larger or less significant. 
We also estimated the descendant of 53W002F-HDR in local. 
The galaxy overdensity of the spherical region at $z$ = 0 which have the same comoving volume and the same rareness probability with 53W002F-HDR is found to be $\sim$ 4, assuming that galaxy bias is unity ($b$ = 1). 
Following \citet{croton05} where clusters are defined as structures whose overdensity of galaxies within 8 h$^{-1}$ Mpc sphere ($\delta_{8}$) is larger than 6.0, the structure whose $\delta_{8}$ equals to 4 (rareness probability $\sim$ 0.9\%) is not classified as a cluster although it is significant high density region. 
On the other hand, it is difficult to find one-to-one correspodence between the regions with similar rareness probability at $z$ = 2.4 and $z$ = 0 because the non-linear effects should be much more seriously considered at lower redshift. 
Therefore we only argue that 53W002-HDR may not evolve to extremely prominent rich cluster at $z$ = 0. 

\subsection{\it The Environmental Dependency of LAEs' Properties}\label{environment-dependence}
In order to discuss the properties of LAEs as a function of the local density environment, we evaluated the LAE overdensities in circular regions with the radius of 2.3$'$ centered at each LAE, which corresponds to the volume of a sphere with radius of 10 Mpc. 
Here we used the 162 LAEs that lie at least 2.3$'$ away from the edge of the image to avoid the edge effect. \\
\begin{figure}[t]
\includegraphics[scale=0.5,angle=270]{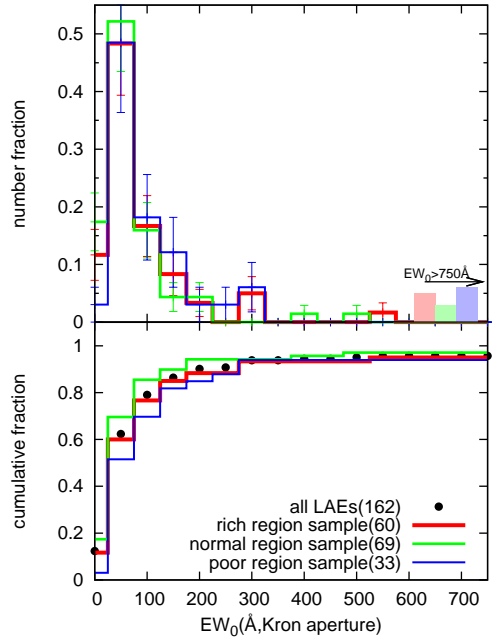}
\caption{The Ly$\alpha$ $EW_0$ distributions in the 53W002 field for the three subsamples divided by their LAE number overdensities: the rich region sample ({\it red histogram}), the normal region sample ({\it green histogram}), and the poor region sample ({\it blue histogram}). The differential ({\it top panel}) and cumulative ({\it bottom panel}) distributions are shown, and the objects out of the $EW_0$ range of the figure are also shown ({\it shaded boxes} in {\it top panel}).\label{fig10}}
\end{figure}
\indent
To search for the dependence of the properties on the environment for the 162 LAEs in the 53W002 field, we divided the samples into the three subsamples with LAE overdensity, $\delta_{LAE}$ = 0.94 and -0.17, which correspond to the rareness probabilities of 12\% and 50\%. 
Hereafter we call the subsample with the LAE overdensity range of -1 $\thicksim$ -0.17, -0.17 $\thicksim$ 0.94 and 0.94 $\thicksim$  as the ``poor region sample'', the ``normal region sample'' and the ``rich region sample'', respectively. 
They contain 60, 69, and 33 LAEs. 
First, we investigated the $EW_0$ distributions. 
Here we used the $EW_0$ measured by the Kron photometry to capture the scattered component of Ly$\alpha$ emission while it actually gives the lower limit, as the $EW_0$ may be affected by possible contamination of the foreground galaxies in the continuum. 
Figure~\ref{fig10} shows the obtained differential ({\it top}) and cumulative ({\it bottom}) distributions of the three subsamples.
There is no notable difference among the three subsamples. 
A KS test shows that every two subsamples have similar distributions with the KS probability larger than 0.25.
We also found that there is no notable difference in the number fraction of LAEs with $EW_0$ larger than 240 \AA : 0.12$\pm$0.04 for the rich region sample, 0.06$\pm$0.03 for the normal region sample, and 0.15$\pm$0.07 for the poor region sample. 
Next, we investigated the distributions of the Ly$\alpha$ luminosities. 
We show the result in Figure~\ref{fig11}. 
The distributions of the three subsamples seem to be in agreement within the error bars. 
There may be a possible weak trend that the rich region sample has a distribution biased to larger Ly$\alpha$ luminosities than others. 
The difference between the rich region sample and others, however, is not significant with KS probability of $\gtrsim$ 0.5. 
It is expected that the Ly$\alpha$ luminosity and $EW_0$ are connected with a mechanism emitting Ly$\alpha$ photons of LAEs. 
In this sense, the Ly$\alpha$ emitting mechanism of LAEs does not depend on their environment (10Mpc scale) at least in the 53W002 field. \\
\begin{figure}[t]
\includegraphics[scale=0.5,angle=270]{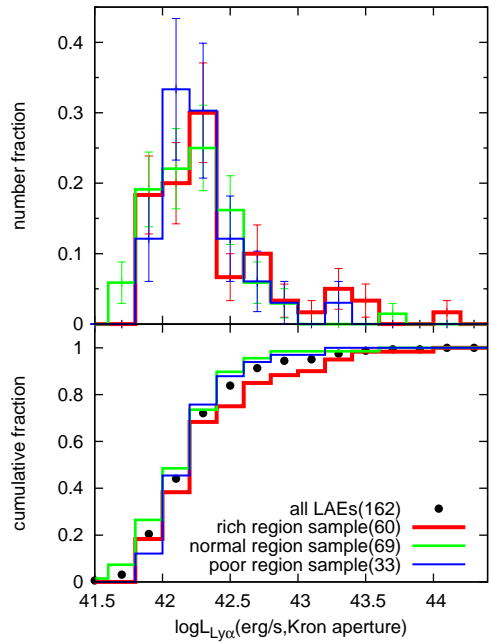}
\caption{The Ly$\alpha$ luminosity distributions in the 53W002 field for the three subsamples, where we use the same relation between color of histograms and subsamples with Fig~\ref{fig10}.\label{fig11}}
\end{figure}
\begin{figure}[t]
\includegraphics[scale=0.55,angle=270]{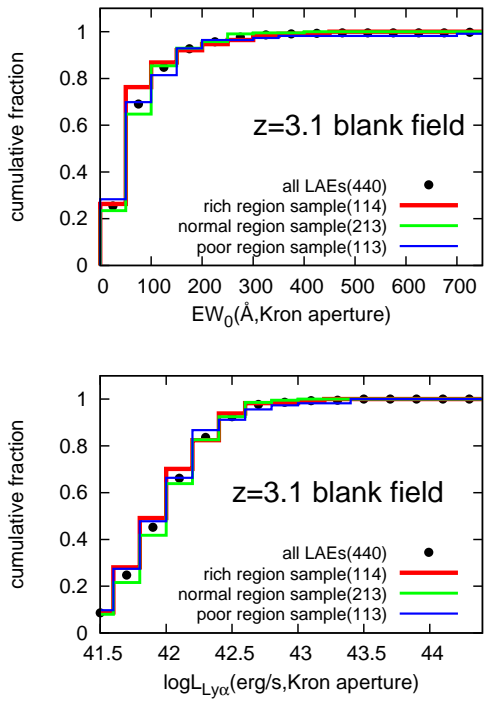}
\caption{The Ly$\alpha$ $EW_0$ ({\it top panels}) and luminosity ({\it bottom panels}) distributions for the {\it z} = 3.1 blank fields (the SXDS field and the SDF field). The LAEs are divided into the three subsamples (see text) and the numbers contained in each subsample are shown in the parentheses of each legend. We rejected the objects near the edge of the images from the original catalog \citep{yamada12}.\label{fig12}}
\end{figure}
\indent
We also investigated the environmental dependency of LAEs in other fields for comparison. 
As supplemental data, 386 LAEs in the SXDS field and 196 LAEs in the SDF field at {\it z} = 3.1 are used from the sample in \citet{yamada12}. 
In their analysis, the SXDS and SDF fields are considered as the blank fields at {\it z} = 3.1 with no significant density excess. 
Before discussing the environmental dependency, we investigated the Ly$\alpha$ $EW_0$ and luminosity distributions of whole sample in the $z$ = 3.1 blank field. 
For the $EW_0$ distribution, we fitted it with an exponentially declining function (Eq.(\ref{eq:2})) and derived the best fit solution of $w_0$ = 64.4 $\pm$ 1.5 \AA. 
This e-folding length at $z$ = 3.1 is significantly larger than $w_0$ = 43.7 $\pm$ 0.43 \AA\ of the 53W002 field at $z$ = 2.4, and this trend is consistent with the discussion in section~\ref{size-ew}. 
On the other hand, we found no notable evolution of Ly$\alpha$ luminosity distribution although the difference of completeness makes the comparison not straightforward. 
In order to search for environmental dependency, we applied almost the same procedure as we did for the LAEs in the 53W002 field: counting the number within the circular region that has the equivalent volume with a sphere with radius of 10 Mpc in comoving scale, rejecting the samples near the edge of the images, dividing the samples into the three subsamples (the rich, normal and poor region sample), and using the measurements based on the Kron photometry. 
To match the {\it z} = 3.1 subsamples with that at {\it z} = 2.4, we used the rareness probability scaled by redshift instead of the observed surface density. 
That is, we adopted the rareness probabilities of 12\% and 50\% as the separators of the subsamples for the total of 440 LAEs in the {\it z} = 3.1 blank fields, where we used the LAE bias of $b_{LAE}$ = 1.9 \citep{guaita10}. 
This corresponds to the LAE number overdensity of 0.85 and -0.13. 
In Figure~\ref{fig12} we summarize the results of our analysis for the {\it z} = 3.1 blank fields. 
KS probabilities for every two subsamples are at least larger than 0.21 in $EW_0$ distributions, and at least larger than 0.34 in L$_{Ly\alpha}$ luminosity distributions. 
There seems to be no dependence of the $EW_0$ or Ly$\alpha$ luminosity distributions on environment in the z = 3.1 blank fields. 
This is consistent with the results in the 53W002 field, and it may be the general property of LAEs. 
We expected that the external environmental effects such as ram pressure, tidal stripping, harassment and so on (which work in timescales of $\sim$ a few Gyr in typical clusters) are not effective because LAEs are thought to be young galaxies (typically have age $\sim$ 10$^7$ yr). 
The results revealed in this section suggest that the intrinsic environmental effects such as galaxy formation bias does not affect the properties of the Ly$\alpha$ emission of LAEs, at least at the scale of 10 Mpc. 
It is also interesting that $EW$ distribution of LAEs changes only along redshift. 
If the $EW$ evolution along redshift is due to evolution of dust amount as claimed in the previous studies \citep{nilsson09}, there should also be environmental dependency since metal enrichment is expected to be proceeded more quickly in high density environment. 
So far there is no definitive answer for this question, but we consider that such environmental information is a key to understand the evolution of LAEs.

\subsection{\it Rich Regions and LABs}
About the four LABs in the 53W002 field we found that they have almost the same rareness probabilities: No.18 and No.19 have 7.7\%, and 53W002 and LAB4 have 10.2\%. 
This clearly supports the view that LAB is a population biased to high density environment \citep{steidel00,matsuda04,prescott08}. 
Interestingly the rareness probabilities of the LABs in the 53W002 field imply that they are not biased to the most dense region but more located at moderate density regions such as the rim of 53W002F-HDR. 
Especially for No.18 and No.19, we consider that galaxy - galaxy mergers may have triggered their AGN activity and induced their morphological features like tail or jet because the merging mechanism becomes effective in outer regions of clusters of galaxies at least in the local universe \citep{treu03}. 
The {\it HST} images of No.18, which shows the double component having different colors and Ly$\alpha$ brightness in the core unresolved in our images, strengthen the merging hypothesis. 
No.19, which lies at the distance of only 80 kpc (physical) from No.18, can be considered as to interact with No.18. 
Unlike the other three, LAB4 doesn't have a strong continuum counterpart such as a QSO, AGN and radio galaxies and looks filamentary. 
Besides, only LAB4 lives far outside 53W002F-HDR. 
It lies with a little offset from the local density peak as the other threes did, and possibly interacts with the two blobby LAEs located in the north of LAB4 (see Figure~\ref{fig7}). 
We speculate that the merger scenario may play a key role in the formation of LABs, but it is needed to confirm the trend that LABs have moderate small rareness probabilities using much larger samples in order to obtain the conclusion. 

\section{CONCLUSION}
We conducted the deep and wide-field imaging of the 53W002 field. \\
\indent
From the spatial distribution of the 204 LAEs in the 53W002 field at {\it z} = 2.4 we identified the overdense region named 53W002F-HDR, where \objectname{53W002} lies at the rim. 
It is found that 53W002F-HDR is a moderately rich region in the {\it z} = 2.4 universe where such overdense structure exist with the rareness probability of $\sim$ 0.9\%. 
The results of the analysis to search for environmental dependency of the properties of LAEs (i.e. Ly$\alpha$ luminosity and $EW_0$) shows that the physical process of emitting Ly$\alpha$ photons of LAEs does not depend on the local density environment at least for 10 Mpc scale. 
This is in contrast to the fact that the $EW_0$ distribution changes along redshift. 
On the other hand, the four LABs (53W002, No.18, No.19 and LAB4) are clearly biased to a moderately rich region such as the rims of the density peaks. 
This trend may give some hints about the origin of the morphological uniqueness of the LABs.

\acknowledgments
We thank the staff of the Subaru Telescope for their assistance. 
This research is supported in part by the Grant-in-Aid 20450224 for Scientific Research of the Ministry of Education, Science, Culture, and Sports in Japan. 



{\it Facilities:} \facility{National Astronomical Observatory of Japan (NAOJ)}.

\setcounter{table}{0}
\begin{deluxetable}{cccccccccll}
\tabletypesize{\scriptsize}
\rotate
\tablecaption{Properties of The LAEs in The 53W002 Field\label{tb1}}
\tablewidth{0pt}
\tablehead{
\colhead{53W002F-} & \colhead{R.A.} & \colhead{Dec.} & \colhead{$EW_{0}$ \tablenotemark{a}} & \colhead{$EW_{0}$ \tablenotemark{b}} & \colhead{log L$_{Ly\alpha}$ \tablenotemark{a}} & \colhead{log L$_{Ly\alpha}$ \tablenotemark{b}} & \colhead{Local density \tablenotemark{c}} & \colhead{$z_{spec}$} & \colhead{Note} & \colhead{Ref}\\
\colhead{LAE\#} & \colhead{(deg)} & \colhead{(deg)} & \colhead{(2$''$,\AA)} & \colhead{(Kron,\AA)} & \colhead{(2'',erg s$^{-1}$)} & \colhead{(Kron,erg s$^{-1}$)} & \colhead{(arcmin$^{-2}$)} & \colhead{} & \colhead{} & \colhead{}\\
\colhead{(1)} & \colhead{(2)} & \colhead{(3)} & \colhead{(4)} & \colhead{(5)} &
\colhead{(6)} & \colhead{(7)} & \colhead{(8)} & \colhead{(9)} & \colhead{(10)} & \colhead{(11)}
}
\startdata
1 & 258.63076 & 50.16096 & 7940 & 40900 & 42.073 & 42.085 & 0.183 & ... & ... & ...\\ 
2 & 258.42144 & 50.25102 & 5310 & $\infty$ & 42.065 & 42.234 & 0.385 & ... & ... & ...\\ 
3 & 258.57732 & 50.30839 & 4970 & 1500 & 42.064 & 42.011 & 0.774 & ... & ... & ...\\ 
4 & 258.54392 & 50.39660 & 865 & 134 & 42.295 & 43.264 & 0.589 & ... & LAB(newly found) & ...\\ 
5 & 258.54149 & 50.39972 & 723 & 48.3 & 42.32 & 42.671 & 0.589 & ... & ... & ...\\ 
6 & 258.54364 & 50.39938 & 310 & 82.1 & 42.595 & 43.062 & 0.589 & ... & ... & ...\\ 
7 & 258.44731 & 50.45681 & 838 & 11500 & 42.004 & 42.077 & 0.376 & ... & ... & ...\\ 
8 & 258.72317 & 50.32180 & 582 & 97.5 & 42.218 & 42.296 & 0.496 & ... & ... & ...\\ 
9 & 258.97269 & 50.41476 & 500 & 1060 & 42.217 & 42.243 & 0.0764 & ... & ... & ...\\ 
10 & 258.59804 & 50.35524 & 474 & 2630 & 42.104 & 42.135 & 0.561 & ... & ... & ...\\ 
11 & 258.18753 & 50.29742 & 454 & $\infty$ & 42.241 & 42.464 & 0.187 & ... & ... & ...\\ 
12 & 258.78259 & 50.29046 & 426 & 45.7 & 42.223 & 42.261 & 0.63 & ... & ... & ...\\ 
13 & 258.62255 & 50.37131 & 386 & 285 & 42.631 & 42.676 & 0.568 & ... & ... & ...\\ 
14 & 258.55015 & 50.26760 & 379 & 312 & 43.397 & 44.066 & 0.694 & 2.393 & LAB/SMG/AGN(No.18) & 2,3,5,6 \\ 
15 & 258.58318 & 50.32840 & 360 & 203 & 42.48 & 42.808 & 0.705 & ... & ... & ...\\ 
16 & 258.42732 & 50.22930 & 351 & 495 & 42.589 & 42.687 & 0.443 & ... & ... & ...\\ 
17 & 258.53700 & 50.40982 & 347 & 312 & 42.034 & 42.282 & 0.607 & ... & ... & ...\\ 
18 & 258.54341 & 50.28229 & 320 & 995 & 42.054 & 42.142 & 0.789 & ... & ... & ...\\ 
19 & 258.76163 & 50.39229 & 314 & 528 & 42.55 & 42.619 & 0.678 & ... & ... & ...\\ 
20 & 258.40484 & 50.16364 & 314 & 313 & 42.396 & 42.424 & 0.21 & ... & ... & ...\\ 
21 & 258.49734 & 50.12923 & 283 & 966 & 42.024 & 42.115 & 0.21 & ... & ... & ...\\ 
22 & 258.79474 & 50.21872 & 264 & 941 & 42.489 & 42.603 & 0.466 & ... & ... & ...\\ 
23 & 258.44204 & 50.33868 & 242 & 295 & 42.714 & 42.805 & 0.248 & ... & ... & ...\\ 
24 & 258.78261 & 50.07203 & 237 & 338 & 42.13 & 42.246 & 0.155 & ... & ... & ...\\ 
25 & 258.84603 & 50.14729 & 221 & 379 & 42.443 & 42.532 & 0.476 & ... & ... & ...\\ 
26 & 258.54748 & 50.42515 & 216 & 1630 & 42.064 & 42.23 & 0.524 & ... & ... & ...\\ 
27 & 258.22744 & 50.28924 & 216 & 243 & 42.297 & 42.327 & 0.161 & ... & ... & ...\\ 
28 & 258.53390 & 50.32310 & 196 & 148 & 42.229 & 42.211 & 0.592 & ... & ... & ...\\ 
29 & 258.66347 & 50.22008 & 195 & 162 & 42.001 & 41.984 & 0.277 & ... & ... & ...\\ 
30 & 258.74374 & 50.25124 & 193 & 103 & 42.004 & 42.199 & 0.422 & ... & ... & ...\\ 
31 & 258.48470 & 50.27243 & 193 & 31.7 & 42.351 & 42.333 & 0.528 & ... & ... & ...\\ 
32 & 258.88279 & 50.11165 & 188 & 53.3 & 42.382 & 42.481 & 0.371 & ... & ... & ...\\ 
33 & 258.66246 & 50.38647 & 171 & 204 & 42.154 & 42.246 & 0.716 & ... & ... & ...\\ 
34 & 258.53123 & 50.28973 & 162 & 151 & 42.257 & 42.37 & 0.817 & ... & ... & ...\\ 
35 & 258.57848 & 50.44524 & 161 & 171 & 42.415 & 42.526 & 0.322 & ... & ... & ...\\ 
36 & 258.49930 & 50.30915 & 155 & 142 & 42.089 & 42.122 & 0.557 & ... & ... & ...\\ 
37 & 258.20361 & 50.43484 & 148 & 106 & 42.8 & 42.879 & 0.141 & ... & ... & ...\\ 
38 & 258.34411 & 50.34501 & 145 & 137 & 42.499 & 42.556 & 0.188 & ... & ... & ...\\ 
39 & 258.27613 & 50.18828 & 136 & 115 & 42.239 & 42.264 & 0.103 & ... & ... & ...\\ 
40 & 258.91993 & 50.24691 & 132 & 95 & 42.509 & 42.767 & 0.318 & ... & ... & ...\\ 
41 & 258.75740 & 50.36116 & 132 & 24.7 & 42.274 & 42.209 & 0.716 & ... & ... & ...\\ 
42 & 258.69606 & 50.39156 & 131 & 96.8 & 42.039 & 42.108 & 0.754 & ... & ... & ...\\ 
43 & 258.21611 & 50.19650 & 131 & 119 & 42.277 & 42.364 & 0.0995 & ... & ... & ...\\ 
44 & 258.41791 & 50.40316 & 128 & 138 & 42.086 & 42.188 & 0.347 & ... & ... & ...\\ 
45 & 258.57892 & 50.27311 & 125 & 113 & 42.081 & 42.138 & 0.652 & 2.396 & ... & 4 \\ 
46 & 258.93658 & 50.17620 & 125 & 62.7 & 42.295 & 42.418 & 0.204 & ... & ... & ...\\ 
47 & 258.75387 & 50.37258 & 123 & 113 & 42.867 & 42.982 & 0.754 & ... & ... & ...\\ 
48 & 258.45500 & 50.41545 & 123 & 181 & 41.955 & 41.993 & 0.504 & ... & ... & ...\\ 
49 & 258.52829 & 50.12714 & 122 & 216 & 42.24 & 42.372 & 0.203 & ... & ... & ...\\ 
50 & 258.59951 & 50.14995 & 122 & 151 & 42.377 & 42.519 & 0.176 & ... & ... & ...\\ 
51 & 258.81284 & 50.23541 & 120 & 80.2 & 42.416 & 42.513 & 0.478 & ... & ... & ...\\ 
52 & 258.44835 & 50.39939 & 117 & 127 & 41.957 & 42.076 & 0.424 & ... & ... & ...\\ 
53 & 258.84204 & 50.15028 & 113 & 24.6 & 42.359 & 42.301 & 0.484 & ... & ... & ...\\ 
54 & 258.59013 & 50.21903 & 113 & 125 & 42.147 & 42.221 & 0.257 & ... & ... & ...\\ 
55 & 258.24914 & 50.09305 & 109 & 70.7 & 42.691 & 42.739 & 0.259 & ... & ... & ...\\ 
56 & 258.73629 & 50.28981 & 109 & 106 & 42.153 & 42.217 & 0.59 & ... & ... & ...\\ 
57 & 258.87868 & 50.41749 & 106 & 48.2 & 42.218 & 42.223 & 0.123 & ... & ... & ...\\ 
58 & 258.82293 & 50.10331 & 100 & 96.5 & 42.456 & 42.504 & 0.389 & ... & ... & ...\\ 
59 & 258.85402 & 50.30794 & 100 & 80.7 & 42.112 & 42.24 & 0.484 & ... & ... & ...\\ 
60 & 258.24106 & 50.11929 & 98.3 & 204 & 42.237 & 42.413 & 0.287 & ... & ... & ...\\ 
61 & 258.24326 & 50.44970 & 98.2 & 95.6 & 42.079 & 42.163 & 0.233 & ... & ... & ...\\ 
62 & 258.35531 & 50.42865 & 97.8 & 362 & 42.125 & 42.419 & 0.272 & ... & ... & ...\\ 
63 & 258.73028 & 50.15968 & 93.5 & 88.1 & 41.958 & 42.143 & 0.148 & ... & ... & ...\\ 
64 & 258.83270 & 50.19312 & 91.9 & 82.6 & 42.152 & 42.252 & 0.454 & ... & ... & ...\\ 
65 & 258.34519 & 50.33534 & 91.8 & 71.2 & 42.35 & 42.443 & 0.188 & ... & ... & ...\\ 
66 & 258.73498 & 50.37537 & 90.7 & 55.1 & 42.02 & 42.006 & 0.772 & ... & ... & ...\\ 
67 & 258.78771 & 50.34518 & 89.7 & 76.6 & 42.027 & 42.201 & 0.581 & ... & ... & ...\\ 
68 & 258.94266 & 50.21324 & 84.6 & 190 & 42.119 & 42.415 & 0.22 & ... & ... & ...\\ 
69 & 258.19161 & 50.10535 & 84.3 & 105 & 42.633 & 42.767 & 0.225 & ... & ... & ...\\ 
70 & 258.86881 & 50.08927 & 83.3 & 130 & 42.153 & 42.234 & 0.294 & ... & ... & ...\\ 
71 & 258.43415 & 50.25486 & 82.9 & 65.4 & 42.01 & 42.014 & 0.453 & ... & ... & ...\\ 
72 & 258.95865 & 50.14175 & 82.6 & 55.9 & 42.341 & 42.591 & 0.139 & ... & ... & ...\\ 
73 & 258.33629 & 50.43656 & 81.1 & 44.1 & 42.011 & 42.015 & 0.32 & ... & ... & ...\\ 
74 & 258.68497 & 50.41743 & 80.6 & 145 & 42.034 & 42.158 & 0.411 & ... & ... & ...\\ 
75 & 258.15448 & 50.40201 & 78.3 & 80.9 & 41.981 & 42.029 & 0.0888 & ... & ... & ...\\ 
76 & 258.53214 & 50.27693 & 78.2 & 62.9 & 42.086 & 42.249 & 0.74 & ... & ... & ...\\ 
77 & 258.55175 & 50.30537 & 76.8 & 79 & 43.246 & 43.302 & 0.814 & 2.381 & QSO & 5 \\ 
78 & 258.38775 & 50.10565 & 76.5 & 77.5 & 43.303 & 43.354 & 0.211 & ... & ... & ...\\ 
79 & 258.65633 & 50.30749 & 76.3 & 51 & 42.027 & 41.975 & 0.525 & ... & ... & ...\\ 
80 & 258.43633 & 50.17643 & 76.2 & 96.7 & 42.213 & 42.493 & 0.357 & ... & ... & ...\\ 
81 & 258.45001 & 50.24254 & 76.1 & 14.3 & 41.984 & 41.651 & 0.493 & ... & ... & ...\\ 
82 & 258.31564 & 50.43478 & 74.8 & 1240 & 42.152 & 42.418 & 0.331 & ... & ... & ...\\ 
83 & 258.69831 & 50.22212 & 74.2 & 59.9 & 41.962 & 42.03 & 0.25 & ... & ... & ...\\ 
84 & 258.62737 & 50.34087 & 73.9 & 173 & 42.01 & 42.254 & 0.651 & ... & ... & ...\\ 
85 & 258.44133 & 50.45587 & 73.8 & 11.8 & 42.02 & 41.816 & 0.363 & ... & ... & ...\\ 
86 & 258.76886 & 50.36924 & 73.1 & 76 & 42.066 & 42.245 & 0.701 & ... & ... & ...\\ 
87 & 258.42478 & 50.44801 & 72.6 & 49 & 42.325 & 42.433 & 0.371 & ... & ... & ...\\ 
88 & 258.28013 & 50.43421 & 71.7 & 70.6 & 42.195 & 42.209 & 0.285 & ... & ... & ...\\ 
89 & 258.74664 & 50.30547 & 71.6 & 95.8 & 41.967 & 42.018 & 0.602 & ... & ... & ...\\ 
90 & 258.48008 & 50.37427 & 70.1 & 94.7 & 42.032 & 42.158 & 0.356 & ... & ... & ...\\ 
91 & 258.87704 & 50.31148 & 68.8 & 9.19 & 42.018 & 41.796 & 0.478 & ... & ... & ...\\ 
92 & 258.63292 & 50.37851 & 68.1 & 22.9 & 42.157 & 42.012 & 0.573 & ... & ... & ...\\ 
93 & 258.79403 & 50.28602 & 67.5 & 182 & 41.901 & 42.022 & 0.566 & ... & ... & ...\\ 
94 & 258.76861 & 50.45149 & 66.7 & 78.3 & 41.974 & 42.103 & 0.177 & ... & ... & ...\\ 
95 & 258.47060 & 50.43591 & 66.4 & 79.2 & 41.906 & 42.029 & 0.477 & ... & ... & ...\\ 
96 & 258.74194 & 50.12380 & 66.3 & 78 & 41.937 & 42.253 & 0.159 & ... & ... & ...\\ 
97 & 258.38649 & 50.09153 & 65.3 & 83.2 & 42.152 & 42.281 & 0.24 & ... & ... & ...\\ 
98 & 258.50209 & 50.21177 & 65 & 73.9 & 42.018 & 42.113 & 0.287 & ... & ... & ...\\ 
99 & 258.60530 & 50.31925 & 64.3 & 49.3 & 41.903 & 41.955 & 0.752 & ... & ... & ...\\ 
100 & 258.44428 & 50.33453 & 62.9 & 63 & 41.933 & 41.95 & 0.248 & ... & ... & ...\\ 
101 & 258.44827 & 50.24072 & 62.7 & 79 & 42.059 & 42.232 & 0.496 & ... & ... & ...\\ 
102 & 258.66369 & 50.31116 & 62.1 & 70.5 & 42.117 & 42.161 & 0.507 & ... & ... & ...\\ 
103 & 258.85948 & 50.06382 & 62.1 & 38.2 & 42.008 & 42.126 & 0.18 & ... & ... & ...\\ 
104 & 258.79800 & 50.24555 & 61.6 & 66.1 & 41.908 & 41.874 & 0.489 & ... & ... & ...\\ 
105 & 258.89831 & 50.33377 & 61.3 & 56.5 & 42.102 & 42.269 & 0.328 & ... & ... & ...\\ 
106 & 258.83134 & 50.36156 & 60.6 & 76.8 & 42.211 & 42.321 & 0.431 & ... & ... & ...\\ 
107 & 258.88995 & 50.26075 & 60.1 & 32.6 & 42.022 & 42.048 & 0.343 & ... & ... & ...\\ 
108 & 258.45175 & 50.24364 & 59.9 & 37.2 & 42.311 & 42.5 & 0.493 & ... & ... & ...\\ 
109 & 258.59660 & 50.24556 & 59.3 & 48.5 & 42.13 & 42.283 & 0.431 & ... & ... & ...\\ 
110 & 258.83743 & 50.35944 & 57.4 & 65 & 42.308 & 42.385 & 0.406 & ... & ... & ...\\ 
111 & 258.90913 & 50.10798 & 56.7 & 48.9 & 41.945 & 41.995 & 0.291 & ... & ... & ...\\ 
112 & 258.80991 & 50.35190 & 56.2 & 63.1 & 42.488 & 42.593 & 0.548 & ... & ... & ...\\ 
113 & 258.58895 & 50.40452 & 56.1 & 57.5 & 41.857 & 41.844 & 0.484 & ... & ... & ...\\ 
114 & 258.92575 & 50.26044 & 55.6 & 59.8 & 41.981 & 42.083 & 0.281 & ... & ... & ...\\ 
115 & 258.83696 & 50.22602 & 55.4 & 48.2 & 41.896 & 41.932 & 0.429 & ... & ... & ...\\ 
116 & 258.81953 & 50.28111 & 53.6 & 94.3 & 42.334 & 42.577 & 0.448 & ... & ... & ...\\ 
117 & 258.49476 & 50.20933 & 53.3 & 67.8 & 42.134 & 42.317 & 0.312 & ... & ... & ...\\ 
118 & 258.31804 & 50.44735 & 52.9 & 43.2 & 42.015 & 41.991 & 0.302 & ... & ... & ...\\ 
119 & 258.31281 & 50.39478 & 52.7 & 52.8 & 41.854 & 41.845 & 0.172 & ... & ... & ...\\ 
120 & 258.27443 & 50.08565 & 52.7 & 32.6 & 42.01 & 41.992 & 0.19 & ... & ... & ...\\ 
121 & 258.80919 & 50.09795 & 52.5 & 44.7 & 42.714 & 42.783 & 0.34 & ... & ... & ...\\ 
122 & 258.81510 & 50.20662 & 52.3 & 75 & 41.994 & 42.156 & 0.472 & ... & ... & ...\\ 
123 & 258.49573 & 50.41455 & 52.2 & 49.5 & 41.922 & 41.912 & 0.557 & ... & ... & ...\\ 
124 & 258.26555 & 50.35800 & 50.8 & 44.3 & 41.965 & 42.016 & 0.146 & ... & ... & ...\\ 
125 & 258.77224 & 50.29308 & 50.7 & 51.1 & 42.254 & 42.352 & 0.644 & ... & ... & ...\\ 
126 & 258.46599 & 50.16240 & 50.5 & 58.6 & 42.616 & 42.754 & 0.361 & ... & ... & ...\\ 
127 & 258.23890 & 50.14443 & 49.8 & 74.1 & 41.984 & 42.149 & 0.175 & ... & ... & ...\\ 
128 & 258.68647 & 50.29608 & 49.3 & 51.1 & 43.663 & 43.725 & 0.452 & 2.393 & QSO & 5 \\ 
129 & 258.62099 & 50.26584 & 48.9 & 11.2 & 41.914 & 41.691 & 0.432 & ... & ... & ...\\ 
130 & 258.51501 & 50.27515 & 48.4 & 28.9 & 42.25 & 42.259 & 0.679 & ... & ... & ...\\ 
131 & 258.66634 & 50.38061 & 48.4 & 53.3 & 42.707 & 42.793 & 0.726 & ... & ... & ...\\ 
132 & 258.76418 & 50.28445 & 48 & 49.2 & 43.536 & 43.591 & 0.632 & ... & ... & ...\\ 
133 & 258.86771 & 50.17787 & 48 & 48.7 & 42.359 & 42.593 & 0.384 & ... & ... & ...\\ 
134 & 258.90682 & 50.11938 & 47.8 & 50.6 & 42.358 & 42.395 & 0.297 & ... & ... & ...\\ 
135 & 258.76462 & 50.23466 & 47.6 & 46.1 & 42.557 & 42.768 & 0.408 & ... & ... & ...\\ 
136 & 258.55367 & 50.32163 & 47.6 & 59.8 & 41.952 & 42.237 & 0.709 & ... & ... & ...\\ 
137 & 258.57258 & 50.44182 & 47.4 & 77.9 & 42.011 & 42.442 & 0.342 & ... & ... & ...\\ 
138 & 258.74577 & 50.41784 & 46.9 & 15.6 & 42.313 & 42.291 & 0.451 & ... & ... & ...\\ 
139 & 258.39931 & 50.08618 & 46.7 & 49.2 & 41.859 & 41.88 & 0.239 & ... & ... & ...\\ 
140 & 258.63872 & 50.20249 & 46.5 & 34.5 & 42.062 & 42.062 & 0.227 & ... & ... & ...\\ 
141 & 258.91421 & 50.22424 & 46.3 & 40.3 & 41.849 & 41.911 & 0.3 & ... & ... & ...\\ 
142 & 258.24874 & 50.10857 & 46.1 & 61 & 42.077 & 42.221 & 0.294 & ... & ... & ...\\ 
143 & 258.82584 & 50.12873 & 45.9 & 24.3 & 41.908 & 41.863 & 0.487 & ... & ... & ...\\ 
144 & 258.18393 & 50.10280 & 44.9 & 105 & 41.956 & 42.229 & 0.196 & ... & ... & ...\\ 
145 & 258.56131 & 50.25854 & 44.5 & 55.4 & 43.02 & 43.398 & 0.567 & 2.390 & LAB/radio(53W002) & 1 \\ 
146 & 258.62468 & 50.24577 & 44.3 & 55.8 & 41.97 & 42.081 & 0.388 & ... & ... & ...\\ 
147 & 258.82171 & 50.19107 & 44.2 & 68 & 41.867 & 42.08 & 0.443 & ... & ... & ...\\ 
148 & 258.69763 & 50.26738 & 44.2 & 45 & 41.857 & 41.924 & 0.376 & ... & ... & ...\\ 
149 & 258.68439 & 50.38154 & 44 & 65.6 & 41.979 & 42.304 & 0.772 & ... & ... & ...\\ 
150 & 258.67207 & 50.37350 & 43.8 & 33.4 & 42.751 & 42.791 & 0.739 & ... & ... & ...\\ 
151 & 258.75152 & 50.40039 & 43.3 & 47 & 41.943 & 41.947 & 0.611 & ... & ... & ...\\ 
152 & 258.59872 & 50.32251 & 43.2 & 27.9 & 42.274 & 42.459 & 0.747 & ... & ... & ...\\ 
153 & 258.51080 & 50.29783 & 42 & 30 & 41.99 & 42.09 & 0.675 & ... & ... & ...\\ 
154 & 258.19402 & 50.30009 & 41.2 & 33.3 & 41.864 & 41.868 & 0.187 & ... & ... & ...\\ 
155 & 258.83558 & 50.39516 & 41 & 42.1 & 41.869 & 41.94 & 0.319 & ... & ... & ...\\ 
156 & 258.43681 & 50.40042 & 41 & 23.7 & 41.823 & 41.747 & 0.395 & ... & ... & ...\\ 
157 & 258.91729 & 50.30043 & 40.8 & 53.6 & 42.067 & 42.14 & 0.338 & ... & ... & ...\\ 
158 & 258.26870 & 50.33844 & 40.4 & 50.4 & 42.656 & 42.792 & 0.16 & ... & ... & ...\\ 
159 & 258.78601 & 50.40570 & 39.1 & 38.5 & 41.824 & 41.857 & 0.455 & ... & ... & ...\\ 
160 & 258.74111 & 50.36884 & 38.7 & 37.5 & 41.804 & 41.862 & 0.749 & ... & ... & ...\\ 
161 & 258.25919 & 50.46269 & 38.7 & 42.2 & 41.945 & 42.041 & 0.228 & ... & ... & ...\\ 
162 & 258.45812 & 50.18303 & 38.1 & 48.8 & 41.862 & 42.012 & 0.389 & ... & ... & ...\\ 
163 & 258.77423 & 50.43241 & 37.8 & 26.3 & 41.867 & 41.945 & 0.329 & ... & ... & ...\\ 
164 & 258.51679 & 50.29996 & 37.7 & 25.4 & 41.939 & 41.949 & 0.736 & ... & ... & ...\\ 
165 & 258.69815 & 50.36658 & 36.9 & 61 & 41.817 & 42.125 & 0.732 & ... & ... & ...\\ 
166 & 258.65537 & 50.24716 & 36.2 & 40.1 & 42.333 & 42.523 & 0.345 & ... & ... & ...\\ 
167 & 258.39421 & 50.06690 & 36.1 & 42.2 & 41.894 & 42.003 & 0.192 & ... & ... & ...\\ 
168 & 258.54232 & 50.14515 & 35.6 & 29.3 & 42.206 & 42.29 & 0.208 & ... & ... & ...\\ 
169 & 258.45903 & 50.44120 & 35.6 & 34.3 & 43.004 & 43.044 & 0.454 & ... & ... & ...\\ 
170 & 258.83864 & 50.15230 & 35.4 & 32.3 & 41.863 & 41.915 & 0.484 & ... & ... & ...\\ 
171 & 258.49989 & 50.38673 & 35.3 & 46.5 & 42.039 & 42.214 & 0.466 & ... & ... & ...\\ 
172 & 258.68790 & 50.41428 & 35 & 56.2 & 41.888 & 42.208 & 0.487 & ... & ... & ...\\ 
173 & 258.41522 & 50.22304 & 33.7 & 44.5 & 42.53 & 42.838 & 0.384 & ... & ... & ...\\ 
174 & 258.67622 & 50.38493 & 33.5 & 112 & 42.17 & 42.557 & 0.746 & ... & ... & ...\\ 
175 & 258.77025 & 50.25271 & 33.1 & 26.1 & 42.001 & 41.958 & 0.489 & ... & ... & ...\\ 
176 & 258.66970 & 50.14040 & 33.1 & 21.9 & 41.872 & 41.899 & 0.144 & ... & ... & ...\\ 
177 & 258.80325 & 50.39305 & 32.4 & 31.2 & 41.974 & 42.03 & 0.476 & ... & ... & ...\\ 
178 & 258.47835 & 50.18144 & 31.1 & 76.2 & 41.934 & 42.206 & 0.36 & ... & ... & ...\\ 
179 & 258.61658 & 50.31336 & 31 & 34.5 & 41.824 & 41.899 & 0.721 & ... & ... & ...\\ 
180 & 258.88997 & 50.30360 & 30.9 & 42.7 & 42.411 & 42.678 & 0.43 & ... & ... & ...\\ 
181 & 258.69585 & 50.39219 & 30.5 & 45.9 & 41.977 & 42.209 & 0.754 & ... & ... & ...\\ 
182 & 258.76024 & 50.30220 & 30 & 24.5 & 41.827 & 41.835 & 0.641 & ... & ... & ...\\ 
183 & 258.62920 & 50.33774 & 29.4 & 26.5 & 41.812 & 41.902 & 0.666 & ... & ... & ...\\ 
184 & 258.74899 & 50.36302 & 29.4 & 22.8 & 41.851 & 41.873 & 0.719 & ... & ... & ...\\ 
185 & 258.44206 & 50.15688 & 29.2 & 1.4 & 41.906 & 41.009 & 0.33 & ... & ... & ...\\ 
186 & 258.95391 & 50.34714 & 28.6 & 22.5 & 41.914 & 41.919 & 0.133 & ... & ... & ...\\ 
187 & 258.89764 & 50.19231 & 28.4 & 29.7 & 42.006 & 42.068 & 0.295 & ... & ... & ...\\ 
188 & 258.61821 & 50.30841 & 28.3 & 45.9 & 42.246 & 42.485 & 0.663 & ... & ... & ...\\ 
189 & 258.39907 & 50.31853 & 28 & 48 & 42.249 & 42.61 & 0.186 & ... & ... & ...\\ 
190 & 258.72235 & 50.30051 & 28 & 39.1 & 42.128 & 42.314 & 0.564 & ... & ... & ...\\ 
191 & 258.80768 & 50.13504 & 28 & 39.3 & 42.662 & 42.878 & 0.443 & ... & ... & ...\\ 
192 & 258.50710 & 50.43131 & 27.8 & 29.2 & 42.043 & 42.159 & 0.478 & ... & ... & ...\\ 
193 & 258.61556 & 50.33513 & 27.5 & 20.7 & 42.343 & 42.457 & 0.701 & ... & ... & ...\\ 
194 & 258.54895 & 50.38175 & 27.4 & 20.6 & 41.803 & 41.872 & 0.48 & ... & ... & ...\\ 
195 & 258.52324 & 50.31402 & 26.5 & 23.5 & 41.848 & 41.92 & 0.677 & ... & ... & ...\\ 
196 & 258.91109 & 50.31318 & 26.3 & 38.1 & 42.101 & 42.343 & 0.383 & ... & ... & ...\\ 
197 & 258.78178 & 50.32572 & 26.2 & 14.8 & 41.96 & 42.011 & 0.562 & ... & ... & ...\\ 
198 & 258.84587 & 50.32118 & 26.1 & -3.5 & 41.958 & NAN & 0.489 & ... & ... & ...\\ 
199 & 258.58729 & 50.41828 & 26 & 25.3 & 41.979 & 42.106 & 0.437 & ... & ... & ...\\ 
200 & 258.86927 & 50.24480 & 25.9 & 25.5 & 42.17 & 42.202 & 0.355 & ... & ... & ...\\ 
201 & 258.54722 & 50.26959 & 25.9 & 34.4 & 43.418 & 43.589 & 0.694 & 2.397 & LAB/QSO(No.19) & 2,3,5 \\ 
202 & 258.83695 & 50.32071 & 25.6 & 14.8 & 41.939 & 42.023 & 0.501 & ... & ... & ...\\ 
203 & 258.83118 & 50.13003 & 25.5 & 23 & 41.809 & 41.803 & 0.487 & ... & ... & ...\\ 
204 & 258.79140 & 50.19397 & 25.4 & 39.7 & 41.885 & 42.071 & 0.393 & ... & ... & ...
\enddata
\tablenotetext{a}{The fourth and sixth rows are based on the fixed aperture (2$''$) photometry.}
\tablenotetext{b}{The fifth and seventh rows are based on the Kron photometry.}
\tablenotetext{c}{The local densities are estimated by smoothing the LAE number density with Gaussian kernel with $\sigma$=1$'$.5.}
\tablerefs{
(1) Windhorst et al. 1991; (2) Pascarelle et al. 1996a; (3) Pascarelle et al. 1996b; (4) Pascarelle et al. 1998; (5) Keel et al. 1999; (6) Smail et al. 2003.}
\tablecomments{There are a few artificial values in this table. The $EW_0$ (Kron) of 53W002F-LAE2 and LAE11 are replaced by ``$\infty$'' because our $EW_0$ calculation gives the continuum fluxes of negative values for these objects. For LAE198, the log L$_{Ly\alpha}$ (Kron) is described as ``NAN'' since we obtained the negative L$_{Ly\alpha}$ (Kron).}

\end{deluxetable}



\appendix

 \section{THE CALCULATION OF THE EW FOR Ly$\alpha$ EMISSION LINES AT {\it z} = 2.4} \label{EWcalc}
Here we show the calculation method of the $EW$ for Ly$\alpha$ at {\it z}=2.4 from the $NB413$ and $B$-band fluxes. \\
\indent
As the wavelength range of the $NB413$ filter is overlapped with that of the $B$-band (see Figure~\ref{fig1}) the continuum flux and the emission flux mix in the $NB413$ flux and the $B$ flux, respectively. 
The $NB413$ flux and the $B$ flux can be described as follows,
\begin{mathletters}
\begin{eqnarray}
f_{{\nu}NB413} & = & f_{{\nu}C} + f_{{\nu}Ly\alpha} , \label{eq:fnuNB}\\
f_{{\nu}B} & = & f_{{\nu}C} + f_{{\nu}Ly\alpha}{\times}\frac{{\Delta}\nu_{NB413}}{{\Delta}\nu_{B}} .\label{eq:fnuB}
\end{eqnarray}
\end{mathletters}
$f_{{\nu}NB413}$/$f_{{\nu}B}$ is the flux per unit frequency (erg s$^{-1}$ cm$^{-2}$ Hz$^{-1}$) for the photons that enter the $NB413$/$B$-band filter, respectively. 
On the other hand, $f_{{\nu}Ly\alpha}$/$f_{{\nu}C}$ is the flux per unit frequency of the Ly$\alpha$ line component/the continuum component at the wavelength of 4140 \AA , respectively. 
${\Delta}\nu_{NB413}$ and ${\Delta}\nu_{B}$ is the width of the $NB413$ and $B$ band filters ($FWHM$) in unit of frequency, respectively. 
Here we assume the flat spectrum of the continuum ($f_{{\nu}C}$ = constant) in the range of the $B$ band because we cannot estimate the continuum slope at $\lambda$ = 4140 \AA\ from only the $B$ and $NB413$ fluxes. \\
From (\ref{eq:fnuNB}) and (\ref{eq:fnuB}),
\begin{mathletters}
\begin{eqnarray}
f_{{\nu}Ly\alpha} & = & \frac{{\Delta}\nu_B}{{\Delta}\nu_{B} - {\Delta}\nu_{NB413}}(f_{{\nu}NB413} - f_{{\nu}B}) , \label{eq:fnuLYA}\\
f_{{\nu}C} & = & \biggl(1 - \frac{{\Delta}\nu_{B}}{{\Delta}\nu_{B} - {\Delta}\nu_{NB413}}\biggr)f_{{\nu}NB413} + \frac{{\Delta}\nu_{B}}{{\Delta}\nu_{B} - {\Delta}\nu_{NB413}}f_{{\nu}B} \label{eq:fnuC}.
\end{eqnarray}
\end{mathletters}
These fluxes per unit frequency lead the several physical quantities of the Ly$\alpha$ emission and the continuum. 
For the Ly$\alpha$ emission, the flux (erg s$^{-1}$ cm$^{-2}$) is
\begin{equation}
F_{Ly\alpha} = f_{{\nu}\alpha} \times {\Delta}\nu_{NB413} = \frac{{\Delta}\nu_B {\Delta}\nu_{NB413}}{{\Delta}\nu_{B} - {\Delta}\nu_{NB413}}(f_{{\nu}NB413} - f_{{\nu}B}) \label{eq:FLYA},
\end{equation}
and the luminosity (erg s$^{-1}$) is\\
\begin{equation}
L_{Ly\alpha} = 4{\pi}d_{L}^{2}F_{Ly\alpha} \label{eq:LLYA},
\end{equation}
where $d_{L}$ is the luminosity distance at {\it z} = 2.4; 19,830 Mpc in the cosmology we adopted. \\
For the continuum, the flux per unit wavelength is\\
\begin{equation}
f_{{\lambda}C} = \frac{c}{{\lambda}^2}f_{{\nu}C} \label{eq:flamC},
\end{equation}
where $\lambda$ = 4140 \AA. \\
Using (\ref{eq:fnuC}), (\ref{eq:FLYA}), and (\ref{eq:flamC}), the $EW$ in rest frame can be calculated as
\begin{equation}
EW_{0} = \frac{F_{Ly\alpha}}{f_{{\lambda}C}} \times \frac{1}{(1+z)} \label{eq:EWrest}.
\end{equation}

\section{THE MAKING OF THE Ly$\alpha$ IMAGE AND THE CONTINUUM IMAGE}\label{mkLYAimage}
In order to make the Ly$\alpha$ image and the continuum image, we should subtract the continuum component from $NB413$ image or the Ly$\alpha$ line component from the $B$ image. 
It is possible to do so using the $NB413$ image and the $B$ image, but careful treatment is needed because the quality of the final image is sensitive to the number of the image operation. \\
The relation between the magnitude and the count in the each image is 
\begin{mathletters}
\begin{eqnarray}
m_{NB413} & = & -2.5 \log(COUNT_{NB413}) + Z_{NB413} ,\label{eq:NmagCOUNT}\\
m_{B} & = & -2.5 \log(COUNT_{B}) + Z_{B} \label{eq:BmagCOUNT},
\end{eqnarray}
\end{mathletters}
where $Z$ is the zero point magnitude of each image; 31.73 for the $NB413$ image and 33.85 for the $B$ image. \\
From these the fluxes per unit frequency can be described as follows,
\begin{mathletters}
\begin{eqnarray}
f_{{\nu}NB413} & = & COUNT_{NB413} \times 10^{-0.4(Z_{NB413} + 48.6)} ,\label{eq:NfnuCOUNT}\\
f_{{\nu}B} & = & COUNT_{B} \times 10^{-0.4(Z_{B} + 48.6)} .\label{eq:BfnuCOUNT}
\end{eqnarray}
\end{mathletters}
From (\ref{eq:fnuLYA}) the Ly$\alpha$ flux per unit frequency is described as
\begin{equation}
f_{{\nu}Ly\alpha} = \frac{{\Delta}\nu_{B}}{{\Delta}\nu_{B} - {\Delta}\nu_{NB413}} \times 10^{-0.4(Z_{B}+48.6)} \bigl\{10^{0.4(Z_{B}-Z_{NB413})} COUNT_{NB413} - COUNT_{B}\bigr\} ,\label{eq:fnuLYAcountBNB}
\end{equation}
and the equation of the Ly$\alpha$ magnitude is 
\begin{eqnarray}
m_{Ly\alpha}&=&-2.5\log(f_{{\nu}Ly\alpha}) -48.6 \nonumber \\
           &=&-2.5\log\bigl\{10^{0.4(Z_{B}-Z_{NB413})} COUNT_{NB413} - COUNT_{B}\bigr\} + Z_{B} + 2.5\log\biggl(1 - \frac{{\Delta}\nu_{NB413}}{{\Delta}\nu_{B}}\biggr) .\nonumber\\\label{eq:mLYAcountBNB}
\end{eqnarray}
Therefore if we take the Ly$\alpha$ image and the zero point of the image as
\begin{mathletters}
\begin{eqnarray}
{\rm Ly{\alpha}\_image} & = & 10^{0.4(Z_{B}-Z_{NB413})} \times {\rm NB413\_image} - {\rm B\_image}\nonumber\\ 
                 & = & 7.047 \times {\rm NB413\_image} - {\rm B\_image} ,\label{eq:LYAimage}\\
Z_{Ly\alpha} & = & Z_{B} + 2.5\log\biggl(1 - \frac{{\Delta}\nu_{NB413}}{{\Delta}\nu_{B}}\biggr) \label{eq:LYAzero},
\end{eqnarray}
\end{mathletters}
(\ref{eq:mLYAcountBNB}) becomes \\
\begin{equation}
m_{Ly\alpha} = -2.5\log(COUNT_{Ly\alpha}) + Z_{Ly\alpha} .\label{eq:mLYAcountLYA}
\end{equation}
So we made the Ly$\alpha$ image as (\ref{eq:LYAimage}). \\
We made the continuum image in the same way. 
From (\ref{eq:fnuC}), (\ref{eq:NfnuCOUNT}), and (\ref{eq:BfnuCOUNT}),
\begin{equation}
f_{{\nu}C} = \frac{{\Delta}\nu_{NB413}}{{\Delta}\nu_{B} - {\Delta}\nu_{NB413}} \times 10^{-0.4(Z_{NB413}+48.6)} \biggl\{\frac{{\Delta}\nu_{B}}{{\Delta}\nu_{NB413}}10^{0.4(Z_{NB413}-Z_{B})} COUNT_{B} - COUNT_{NB413}\biggr\} ,\nonumber\\ \label{eq:fnuCcountBNB}
\end{equation}
and
\begin{equation}
m_{C} = -2.5\log\biggl\{\frac{{\Delta}\nu_{B}}{{\Delta}\nu_{NB413}}10^{0.4(Z_{NB413}-Z_{B})} COUNT_{B} - COUNT_{NB413}\biggr\} + Z_{NB413} + 2.5\log\biggl(\frac{{\Delta}\nu_{B}}{{\Delta}\nu_{NB413}} - 1\biggr) .\nonumber\\\label{eq:mCcountBNB}
\end{equation}
Therefore we made the continuum image and the zero point of the image as
\begin{mathletters}
\begin{eqnarray}
{\rm continuum\_image} & = & \frac{{\Delta}\nu_{B}}{{\Delta}\nu_{NB413}}10^{0.4(Z_{NB413}-Z_{B})} \times {\rm B\_image} - {\rm NB413\_image}\nonumber\\ 
                 & = & 1.683 \times {\rm B\_image} - {\rm NB413\_image} ,\label{eq:Cimage}\\
Z_{continuum} & = & Z_{NB413} + 2.5\log\biggl(\frac{{\Delta}\nu_{B}}{{\Delta}\nu_{NB413}} - 1\biggr) . \label{eq:Czero}
\end{eqnarray}
\end{mathletters}





\end{document}